\documentclass{ws-ijmpb}

\usepackage{array}

\usepackage{booktabs}

\newcommand{\vc}[1]{\boldsymbol{#1}}

\UseRawInputEncoding
\begin{document}


%
\catchline{}{}{}{}{}
%

\title{Towards Kitaev Spin Liquid in 3$\boldsymbol{d}$ Transition Metal Compounds}

\author{Huimei Liu}

\address{Max Planck Institute for Solid State Research,
Heisenbergstrasse 1, D-70569 Stuttgart, Germany\\
H.Liu@fkf.mpg.de }

\maketitle

\begin{history}
\received{Day Month Year}
\revised{Day Month Year}
\end{history}

\begin{abstract}
This paper reviews the current progress on searching the Kitaev spin liquid state in 3$d$ electron systems.
Honeycomb cobaltates were recently proposed as promising candidates to realize the Kitaev spin liquid state,
due to the more localized wave functions of $3d$ ions compared with that of $4d$ and $5d$ ions,
and also the easy tunability of the exchange Hamiltonian in favor of Kitaev interaction.
Several key parameters that have large impacts  on the exchange constants, such as the charge-transfer gap and the trigonal crystal field, are identified and discussed.
Specifically, tuning crystal field effect by means of strain or pressure is emphasized as an efficient phase control method driving the magnetically ordered cobaltates into the spin liquid state.
Experimental results suggesting the existence of strong Kitaev interactions in layered honeycomb cobaltates are discussed.
Finally, the future research directions are briefly outlined.
\end{abstract}


\section{Introduction}

Transition metal compounds with $4d$ or $5d$ ions have become one of the main focus of condensed matter physics recently,
where the spin-orbit coupling (SOC) effect is highlighted. It is believed that the $4d$ and $5d$ systems,
where both correlation physics and non-perturbative SOC physics come into play, could provide a platform
to realize exotic phases of matter such as quantum spin liquids, unconventional
superconductivity, and various topologically nontrivial states.

$3d$ transition metal compounds, where high-$T_c$ superconductivity, colossal magneto-resistance, multiferroics, and exotic spin-charge-orbital orderings
were first discovered in, have been gradually forgotten in the context of SOC related research.
This may be partially due to the common belief among the current generation  of researchers that SOC effects are suppressed in $3d$ systems since the
SOC strength is smaller compared with that in $4d$ or $5d$ ions.
However, one has to keep in mind that the relevance of SOC in a given material is not decided
by the absolute value  of coupling strength alone, but by its comparison with other couplings. As long as the spin-orbit coupling
can overcome the exchange and orbital-lattice interactions, the entanglement of spin and orbital degrees of freedom is essential
while describing the low energy physics in the material.

In fact, late $3d$ transition metal compounds are SOC systems with long history.
For instance, the cobaltates known as  spin-orbit entangled materials, were already well studied in last century.
Strong SOC induced magnetic anisotropy in Co compounds was used by Fert and Gr$\rm \ddot{u}$nberg
to design the Nobel Prize winning giant magnetoresistance (GMR) device which is widely used nowadays.
Therefore, all the exotic SOC-related physics discussed in $4d$ and $5d$ materials must be present and deserves looking for in $3d$ systems.

Very recently, it was theoretically proposed that the ``vintage'' SOC systems $3d$ cobaltates are indeed very promising
candidates to realize the quantum spin liquid state with nontrivial topological properties. After that, there
has been increasing experimental efforts devoted to verify
this proposal.
Motivated by this, we review particularly the recent progress of this direction
and discuss the potential to realize the spin liquid state in $3d$ cobaltates.

\section{The Kitaev honeycomb model}
In 1973, the quantum spin liquid (QSL) state was first proposed by Anderson as the ground state
for nearest neighbor spins $S=1/2$ interacting antiferromagnetically on a triangular lattice systems\cite{And73},
which he referred to as ``resonating valence bond'' state.
In 1987, further interest in QSL has been promoted by the discovery of
high-$T_c$ superconductivity which was suggested arising from doping a QSL\cite{And87,Bas87}.
Ever since, QSL has been one of the most pursued magnetic states.
After decades of tremendous efforts, the conceptual understanding of QSL state has been gradually advanced.
However, the concern of whether the QSL could really occur in real physical system
has not been finally dispelled until 2006 when Kitaev proposed his honeycomb model\cite{Kit06} with an exact solution and a stable gapless QSL ground state; for extensive discussions of this model,
see the recent reviews\cite{Sav17,Her18,Tre17,Win17,Tak19,Mot20,Tom21}.

\begin{figure}
\begin{center}
\includegraphics[width=11cm]{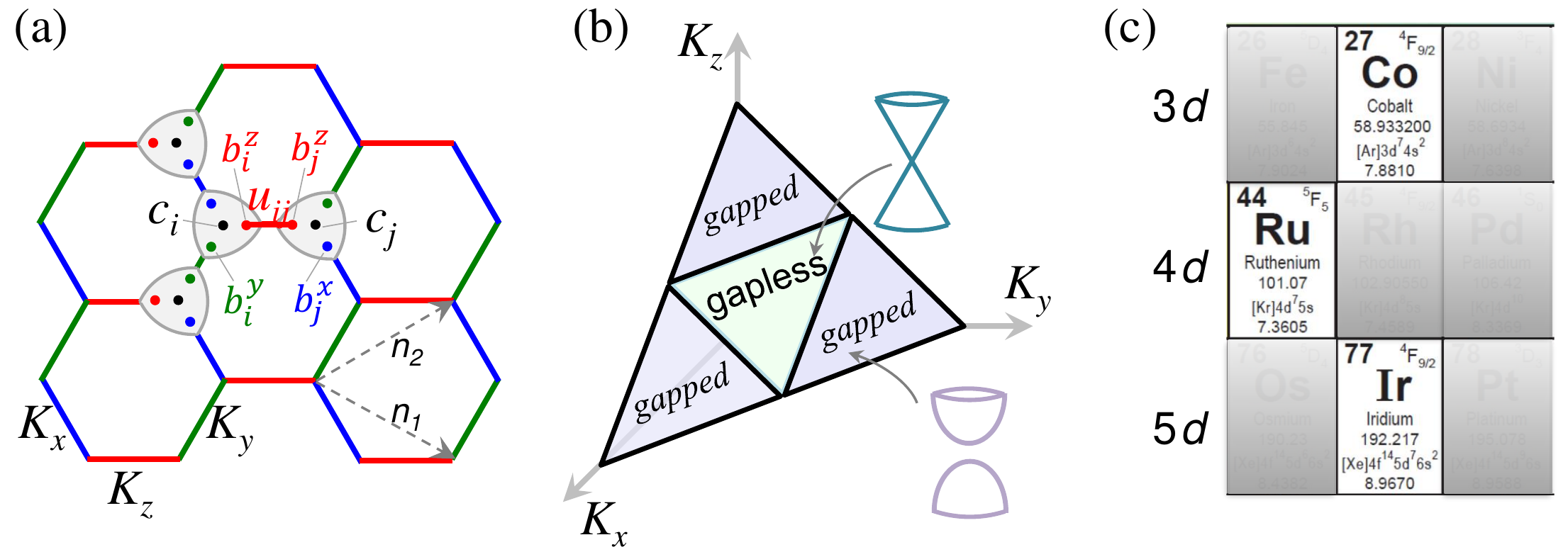}
\caption{(a) The graphic representation of Kitaev honeycomb model with bond-directional couplings $K_x$, $K_y$ and $K_z$.
Four flavors of Majorana fermions are indicated by the black, blue, green and red filled dots. $\vc n_1 = (1/2,\sqrt{3}/2)$ and $\vc n_2 = (-1/2,\sqrt{3}/2)$ are lattice basis vectors for the hexagon.
(b) Phase diagram of the Kitaev model on a plane of $K_x+K_y+K_z=$const.
(c) The Kitaev honeycomb model can be realized in Mott insulators of late transition metal ions with strong spin-orbit coupling, e.g. Co, Ru and Ir.}
\label{fig:str}
\end{center}
\end{figure}

In the Kitaev honeycomb model, the nearest-neighbor (NN) spins $S=1/2$ interact via a simple Ising-type
coupling:
\begin{align}
\mathcal{H} = - \sum_{\gamma} K_{\gamma} S_i^{\gamma} S_j^{\gamma}  \;,
\label{eq:HKi}
\end{align}
where $\gamma \in x,y,z$ indicate the three types of bonds in a honeycomb lattice as shown in Fig.~\ref{fig:str}(a).
It is clear that the Ising axis $\gamma$ is bond-dependent, taking
the mutually orthogonal directions ($x, y, z$) on the three adjacent NN-bonds of
the honeycomb lattice. Having no unique easy-axis and being frustrated, the Ising
spins fail to order and realize instead the quantum spin liquid.

This model can be exactly solved and allows one to precisely describe
the fractionalization of spin degrees of freedom into an emergent Majorana fermion and a $Z_2$ gauge field.
One can define $S_i^{\gamma}=\frac{i}{2} b_i^{\gamma}c_i$, where $b_i^{\gamma}$ and $c_i$ represent
four Majorana modes with the constraint $b_i^x b_i^y b_i^z c_i=1$ to preserve the
two-dimensional Hilbert space and satisfy the algebraic relations for $S=1/2$.
In the Majorana representation, Eq.~\ref{eq:HKi} can be rewritten as:
\begin{align}
\mathcal{H} 
= \frac{i}{4} \sum_{\gamma} K_{\gamma} u_{ij} c_i c_j  \;,
\label{eq:HKi2}
\end{align}
where bond variables $ u_{ij} = i b_i^{\gamma} b_j^{\gamma} $ commute  with each other $[u_{ij}, u_{kl}]=0$ and with the Hamiltonian $[u_{ij}, \mathcal{H}]=0$, and therefore are conserved quantities.
Thus, we have $ u_{ij}= \pm 1$ defining an orthogonal decomposition of the full Fock space,
and the operator $u_{ij}$ in Eq.~\ref{eq:HKi2} can be replaced by numbers.

Without further constraint, there are many selections of $\{ u_{ij} \}$. To fix it, let's first define a plaquette operator:
\begin{equation}
W_p = 2^6 S_1^x S_2^y S_3^z S_4^x S_5^y S_6^z , \ \ \ \ \ \ \ \ \ \
\raisebox{-28pt} {\includegraphics[height=58pt]{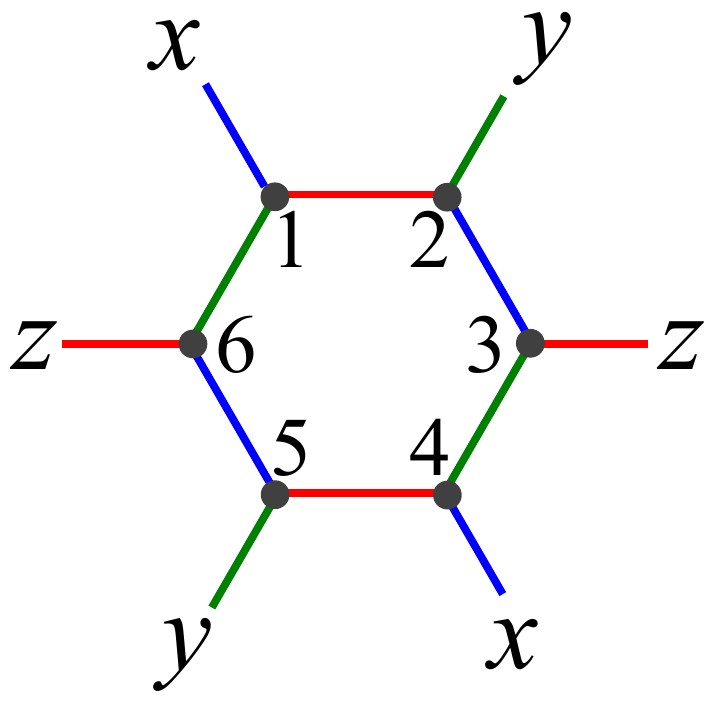}} \;,
\label{eq:pl}
\end{equation}
where the spins and labels follow from the figure next to the equation. 
This $Z_2$ valued operator commutes with the Hamiltonian $[W_p,\mathcal{H}]=0$, i.e. each $W_p=\pm 1$ per hexagon.
Within the physical subspace, the plaquette
operator can be rewritten as:
\begin{align}
W_p= \prod_{\langle ij \rangle\in \partial p} u_{ij} .
\label{eq:HKi2}
\end{align}
According to a theorem by Lieb\cite{Lie94}, the ground
state has no vortices, which is, $W_p=1 ~ \forall p$.
One acceptable selection of $ \{u_{ij} \}$ to satisfy this constraint is $u_{ij}=1$ when
$i$ belongs to the one sublattice and $ u_{ij}=-1$ for $i$ belongs to the other sublattice of the honeycomb lattice.
Then, one can diagonalize the model Eq.~\ref{eq:HKi2} and obtain the following dispersion for $c$ fermion excitations:
\begin{align}
\varepsilon_{\vc k}=\pm |K_x e^{i \vc k \cdot \vc n_1} + K_y e^{i \vc k \cdot \vc n_2} + K_z| .
\end{align}
Here, $\vc n_1$ and $\vc n_2$ are lattice basis vectors shown in Fig.~\ref{fig:str}(a).

Two different phases can be realized while changing the values of $K_{\gamma}$, as shown in the phase diagram of Fig.~\ref{fig:str} (b).
In the gapless phase around the point with equal coupling $K_x = K_y = K_z$,
the fermion spectrum contains two zero-energy Dirac points which will merge and disappear
at the transition to the gapped phase. The latter state is an $Abelian$ topological phases
and can be connected to the Kitaev's toric code model\cite{Kit03}.
The $Z_2$ gapless phase is of particular interest since it can gap out into a massive $non$-$
Abelian$ topological phase by applying a perturbation which breaks the time-reversal symmetry, for instance, magnetic fields.
Within this massive phase,
the spectral Chern number is finite and determines robust chiral modes at the edge
which can be used to perform braiding operations for the fault-tolerant quantum computation.
Thus, the searching of real materials with $K_x = K_y = K_z$ has become more and more popular
 since the proposal of the Kitaev honeycomb model.

With the increasing focus on material realization of this model,
several physical observables have been discussed such as the dynamic spin structure factor\cite{Kno14a} and
Raman response\cite{Kno14b,Nas16}, among which, the very unique signature of chiral Majorana edge modes is the half-integer thermal Hall effect with a quantized Hall
conductivity $\kappa_{xy}/T=\frac{1}{2}\frac{\pi k_B^2}{6 \hbar}$ in the low-temperature limit\cite{Kit06,Nas17}.
There have also been many extended studies of this model such as
the disorder effect\cite{Wil10} and $p$-wave superconductivity induced by doping\cite{Hya12,You12}.

\section{The  Kitaev model in real materials}
To realize the Kitaev model, many schemes have been proposed such as by means of
cold atoms\cite{Dua03,Mic06,Gor13}, organic materials\cite{Yam17a,Yam17b}, superconducting networks\cite{You10}
and magnetic clusters\cite{Wan10}. In this review, we focus on the spin-orbital entangled materials based on transition metal ions shown in Fig.~\ref{fig:str}(c).

When the SOC dominates over the exchange and orbital-lattice interactions,
the orbital moment $\vc L$ remains unquenched and a total angular momentum $\vc J =\vc S+ \vc L$ is formed.
The spin interactions are normally SU(2) invariant since the total spin is conserved
during the electron exchange processes.
On the other hand, the orbital exchange interactions are far more complicated: they are highly frustrated and anisotropic in both real and magnetic spaces\cite{Kha05,Kug82,Kha03}.
Inherited by the ``pseudospins'' $J$ via SOC, the orbital magnetism has become an origin of nontrivial
interactions and exotic phases such as spin-orbit Mott insulator, excitonic magnetism, multipolar magnetism,
quantum spin liquid, and topological phases. The anisotropic and bond-dependent exchange interactions between orbitals
is desired by the Kitaev honeycomb model. Therefore, the key receipt of realizing the Kitaev-type interactions in real materials is to include the
orbital magnetism, which has been certified in $d^5$ systems\cite{Kha05,Jac09,Cha15}.

Along this line, Jackeli and Khaliullin have suggested to realize the Kitaev honeycomb model in $5d^5$ iridates\cite{Jac09}
with pseudospin-1/2 ground state under strong SOC in 2009.
Later on, $4d$ ruthenates\cite{Plu14} were also proposed to host the Kitaev honeycomb model.
To date, quite a number of materials have been proven to host strong bond-directional
interactions, such as Na$_2$IrO$_3$\cite{Cha10,Chu15}, $\alpha$-RuCl$_3$\cite{Ban16,Ban17,Do17,Ban18,Kas18} and so on\cite{Sin12,Tak15,Kit18,Bah19}.
However, instead of forming spin liquid state at low temperature, most of the candidate materials display long range magnetic orders,
caused by the existence of additional exchange couplings such as the Heisenberg interaction
$J$, non-diagonal anisotropy $\Gamma$ and $\Gamma'$ terms within NN bonds, and unavoidable longer range interactions.
In this section, we will explain the origin of the additional exchange interactions and briefly
discuss the difficulties of realizing the pure Kitaev  honeycomb model in 4$d$ and 5$d$ systems.

\begin{figure}
\begin{center}
\includegraphics[width=12.6cm]{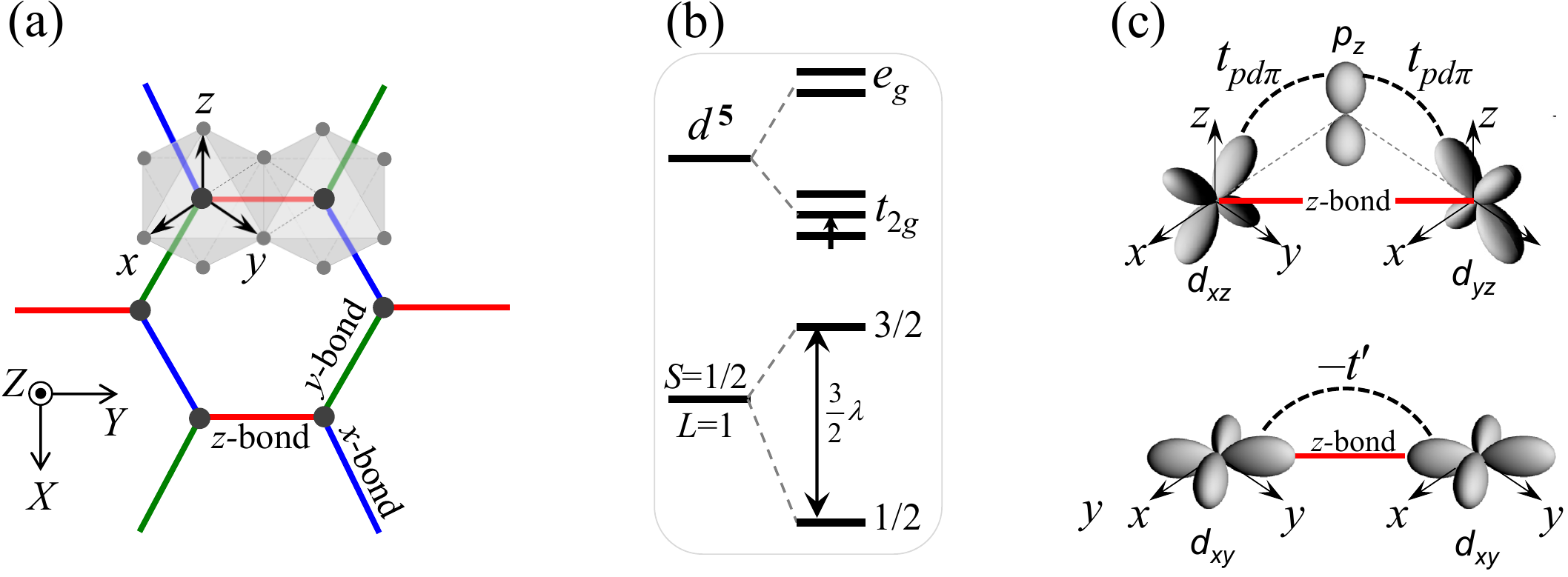}
\caption{(a) Top view of the honeycomb lattice, $x$-, $y$- and $z$-type NN-bonds
are shown in blue, green and red colors,
respectively. Each transition metal ion (black balls) is surrounded by six
ligand anions (grey dots) forming an octahedra.
The definition of global $X$, $Y$, $Z$ and the local cubic $x$, $y$, $z$ axes are shown.
(b) Level structures of the $d^5$ configuration in an octahedral
crystal field without (upper panel) and with (lower panel) SOC in a hole representation.
(c) Hopping processes between $t_{2g}$ orbitals along $z$-type NN-bond for the
ideal 90$^{\circ}$ bonding geometry.}
\label{fig:1}
\end{center}
\end{figure}

The general lattice structure shared by honeycomb iridates and ruthenates is shown in Fig.~\ref{fig:1}(a),
the Ir$^{4+}$ or Ru$^{3+}$ ion is surrounded by six ligand ions forming an octahedron.
The five electrons of Ir$^{4+}$ or Ru$^{3+}$ ion all reside on $t_{2g}$
orbitals due to the strong cubic crystal field. This $t_{2g}^5$ electronic configuration forms a low spin $S=1/2$ state as shown in Fig.~\ref{fig:1}(b).
The threefold orbital degeneracy of this configuration can be described in terms of
an effective angular momentum $L=1$\cite{Abr70} with the following relations:
\begin{alignat}{2}
a& = -\frac{1}{\sqrt{2}}(d_1-d_{-1}),
&\quad \ \ \ \ \ \ \ \ \ \ \ \ \
d_1^{\dag} &= -\frac{1}{\sqrt{2}}(a^{\dag}+ib^{\dag}),
\notag \\
b& = \frac{1}{i\sqrt{2}}(d_1+d_{-1}),
&
d_{-1}^{\dag} &= \frac{1}{\sqrt{2}}(a^{\dag}-ib^{\dag}),
\notag \\
c& = d_0,
&
d_0^{\dag} &= c^{\dag}\;.
\label{eq:ab}
\end{alignat}
Here, the short notations $a= d_{yz}$,  $b= d_{zx}$ and  $c= d_{xy}$ are introduced for convenience,
the indices $0$ and $\pm1$ stand for effective angular momentum
$L_z$ projections [in the quantization axes specified in Fig.~\ref{fig:1}(a)].
Diagonalization of $H_{\lambda}=\lambda \vc L \cdot \vc S$ results in a level structure
which are labeled according to the total angular momentum $J=$1/2 and 3/2, as shown in Fig.~\ref{fig:1}(b).
The ground Kramers doublet hosts the pseudospin-1/2 state with the wavefunctions, written in the basis of $|S_z,L_z \rangle$, read as:
\begin{align}
\Big|\!\pm \frac{1}{2}\Big\rangle  =
\sqrt{\frac{1}{3}} \Big|0,\pm\frac{1}{2} \Big\rangle
\mp \sqrt{\frac{2}{3}} \Big|\pm 1,\mp\frac{1}{2} \Big\rangle.
\end{align}

In an ideal honeycomb lattice, two NN transition metal ions are bridged by two ligand ions with the bonding angle equals
90$^{\circ}$. The hopping between $t_{2g}$
orbitals along the $\gamma=z$-type NN-bonds can be written as\cite{Kha05,Kha04,Nor08,Cha11}:
\begin{align}
\mathcal{H}^z_t = \sum_{\sigma} \left[t(a_{i\sigma}^{\dag}b_{j\sigma}+b_{i\sigma}^{\dag}a_{j\sigma})
-t'c_{i\sigma}^{\dag}c_{j\sigma} + \mathrm{H.c} \right].
\label{eq:tt'}
\end{align}
Here, $\sigma$ is spin index, $t= t_{pd\pi}^2/\Delta_{pd}$ is the hopping amplitude between $a= d_{yz}$ and $b= d_{zx}$ orbitals,
$t'>0$ is the direct overlap between $c= d_{xy}$ orbitals, see Fig.~\ref{fig:1}(c).

With Eq.~\ref{eq:ab}, the above hopping Hamiltonian Eq.~\ref{eq:tt'} can be translated into:
\begin{align}
\mathcal{H}^z_t = \sum_{\sigma}\left[
it(d_{1,\sigma}^{\dag}d_{-\!1,\sigma}-d_{-\!1,\sigma}^{\dag}d_{1,\sigma})_{ij}
-t'(d_{0,\sigma}^{\dag}d_{0,\sigma})_{ij}+\mathrm{H.c}\right].
\label{eq:tt'2}
\end{align}
Even without the detailed calculations, one can have some  idea about the resulting exchange Hamiltonian by
observing Eq.~\ref{eq:tt'2}: the $t$ hopping processes change the total angular momentum by
$\pm 2$ and thus can not connect the pseudospin-1/2 state, unless higher order processes
such as hoppings to $J=3/2$ or $e_g$ states via Hund's coupling are included.
Once the higher order processes are considered, they lead to a ferromagnetic (FM) Kitaev interaction $K$ with the magnitude proportional to $ \frac{J_H}{U}\frac{t^2}{U}$\cite{Jac09}.
On the other hand, the direct hopping process $t'$, which conserves
the total angular momentum, gives rise to antiferromagnetic (AFM) Heisenberg coupling $J \propto \frac{t'^2}{U}$.
Despite of the fact $t'<t$, the Heisenberg coupling $J$ is generally expected to be comparable with
$K$, since the Kitaev interaction is given by higher order effect $\frac{J_H}{U}$.
This may be one of the intrinsic disadvantages of realizing the Kitaev spin liquid (KSL) phase in $d^5$ materials.

\begin{figure}[bt]
\begin{center}
\includegraphics[width=12.6cm]{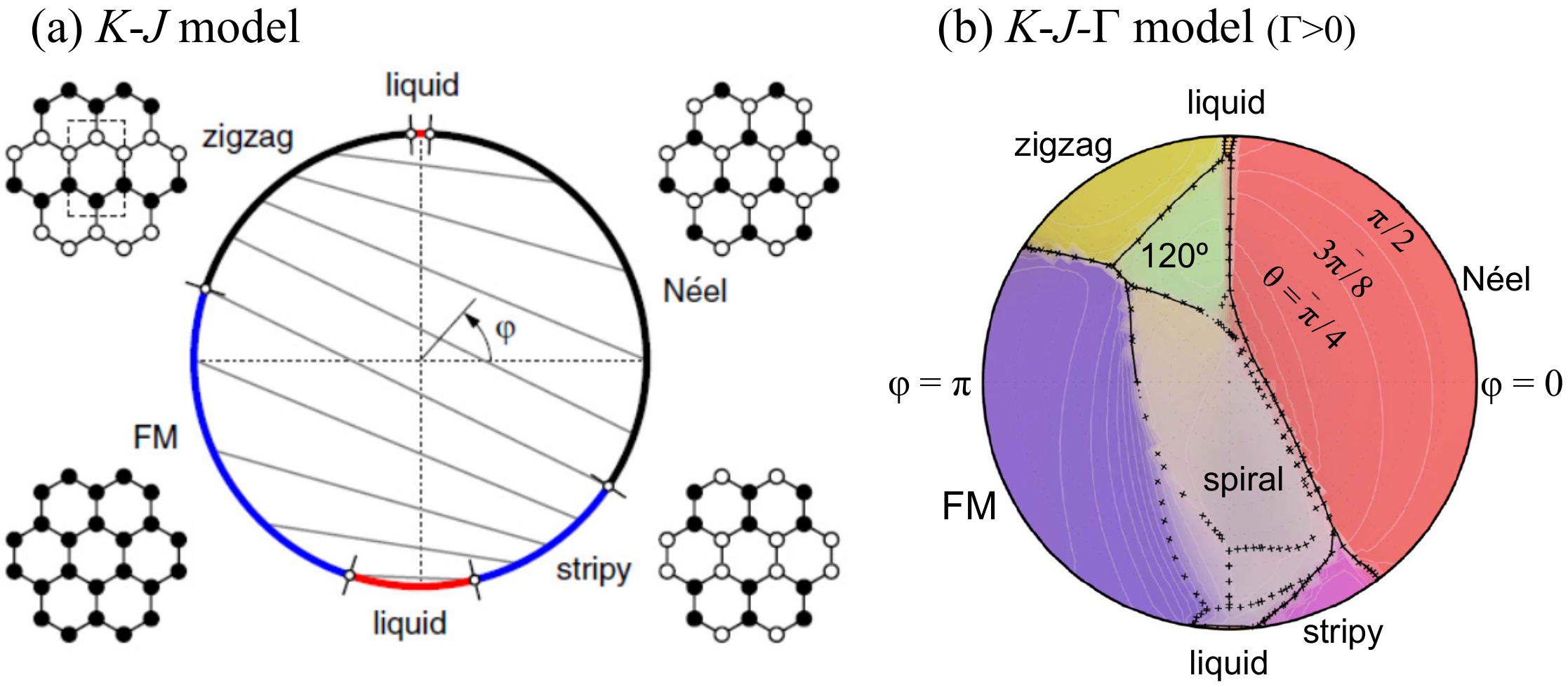}
\caption[]{(a) Phase diagram of the $K$-$J$ model: $\mathcal{H}_{ij}^{z}=K \widetilde{S}_i^z\widetilde{S}_j^z +
J  \widetilde{\vc S}_i \cdot  \widetilde{\vc S}_j=A(2 \sin \varphi \widetilde{S}_i^z\widetilde{S}_j^z +
\cos \varphi  \widetilde{\vc S}_i \cdot  \widetilde{\vc S}_j)$. The phase angle $\varphi$ varies from 0 to $2\pi$ and the corresponding magnetic
phases are indicated in the figure, which is taken from Ref.~\refcite{Cha13}.
(b) Phase diagram of the  $K$-$J$-$\Gamma$ model: $\mathcal{H}_{ij}^z=K \widetilde{S}_i^z\widetilde{S}_j^z +
J  \widetilde{\vc S}_i \cdot  \widetilde{\vc S}_j+
\Gamma (\widetilde{S}_i^x\widetilde{S}_j^y +\widetilde{S}_i^y\widetilde{S}_j^x)
=A[\sin \theta \sin \varphi \widetilde{S}_i^z\widetilde{S}_j^z +
\sin \theta \cos \varphi  \widetilde{\vc S}_i \cdot  \widetilde{\vc S}_j
+\cos \theta (\widetilde{S}_i^x\widetilde{S}_j^y +\widetilde{S}_i^y\widetilde{S}_j^x)]$.
Figure is reproduced from Ref.~\refcite{Rau14} and
$\theta$ varies from 0 to $\pi/2$ corresponds to positive $\Gamma>0$.
See also Ref.~\refcite{Rau14} for phase diagram with negative $\Gamma<0$.}
\label{fig:2}
\end{center}
\end{figure}

Considering that the Heisenberg interaction in real materials can be as large as the Kitaev term,
 Chaloupka $et$ $al.$ have calculated the phase diagram
of the $K$-$J$ model\cite{Cha13} using exact diagonalization (ED) method. The results are shown in Fig.~\ref{fig:2}(a).
The KSL phase can be stabilized for both FM and AFM $K$ when the Heisenberg
interaction is not very strong, which leaves some room for the material search.
In addition, the Heisenberg exchange gives rise to several magnetic ordered states surrounding the liquid phase,
such as the zigzag state corresponds to the cases of Na$_2$IrO$_3$ and $\alpha$-RuCl$_3$. 

Later, Rau $et$ $al.$ have found out that the direct hopping $t'$ process can also induce
a non-diagonal anisotropy often referred to as $\Gamma$ term\cite{Rau14}.
They have calculated the phase diagram of the extended $K$-$J$-$\Gamma$ model
with ED method as presented in Fig.~\ref{fig:2}(b). Compared with Fig.~\ref{fig:2}(a), more ordered states
are introduced by the $\Gamma$ term at the cost of suppressing the liquid state.

By symmetry, the nearest-neighbor exchange Hamiltonian in materials with ideal
honeycomb lattice is of the following general form (for $z$-type of bonds):
\begin{align}
\mathcal{H}_{ij}^{z}=K \widetilde{S}_i^z\widetilde{S}_j^z +
\underbrace{J  \widetilde{\vc S}_i \cdot  \widetilde{\vc S}_j +
\Gamma (\widetilde{S}_i^x\widetilde{S}_j^y +\widetilde{S}_i^y\widetilde{S}_j^x  )}_{\rm direct~hopping~\emph{t}'~processes}
+ \underbrace{\Gamma'(\widetilde{S}_i^x\widetilde{S}_j^z +
\widetilde{S}_i^z\widetilde{S}_j^x+
\widetilde{S}_i^y\widetilde{S}_j^z +
\widetilde{S}_i^z\widetilde{S}_j^y )}_{\rm trigonal~crystal~field}.
\label{eq:H}
\end{align}
The interactions on $x$- and $y$-type NN-bonds can be obtained by cyclic permutations among
$\widetilde{S}_j^x$, $\widetilde{S}_j^y$, and $\widetilde{S}_j^z$
(here we use $\widetilde{S}$ instead of $J$ for pseudospin-1/2 to avoid confusion
between the notations of pseudospins and exchange couplings). This Hamiltonian has very rich and nontrivial symmetry properties,
as discussed in great details in Ref.~\refcite{Cha15}.
In addition to Eq.~\ref{eq:H}, which is referred to as ``the extended Kitaev model'', the full exchange
Hamiltonian has to be also supplemented by longer range couplings\cite{Win16}, which are unavoidable in weakly
localized 5$d$- and 4$d$-electron systems with the spatially extended $d$ wave functions.
With these additional exchange couplings, it is impossible to solve Eq.~\ref{eq:H} exactly.
Luckily, the numerical methods are capable of suggesting that the KSL phase is still stable when the Kitaev interaction
are dominant over the other couplings.

As indicated in Eq.~\ref{eq:H}, the direct hopping $t'$ processes generate the ``non-Kitaev'' $J$ and $\Gamma$ couplings.
Therefore, materials with smaller $t'$, and thus smaller $J$ and $\Gamma$ interactions, are better candidates to realize the exotic KSL phase.
Following this simple logic, it seems that $3d$ transition metal compounds with spatially less extended wave functions
can meet the requirement.
Moreover, the spatially compact wave functions should, in principle, also have smaller contributions to
longer range interactions.

\section{Pseudospin-1/2 ground state in $3d^7$ cobaltates}
The idea of extending the search of the Kitaev materials to $3d$ systems seems straightforward and promising.
However, there is an important question
to be addressed in the first place: is SOC in 3$d$ ions strong enough to support the orbital magnetism prerequisite for the Kitaev model design?

\begin{figure}
\begin{center}
\includegraphics[width=12.6cm]{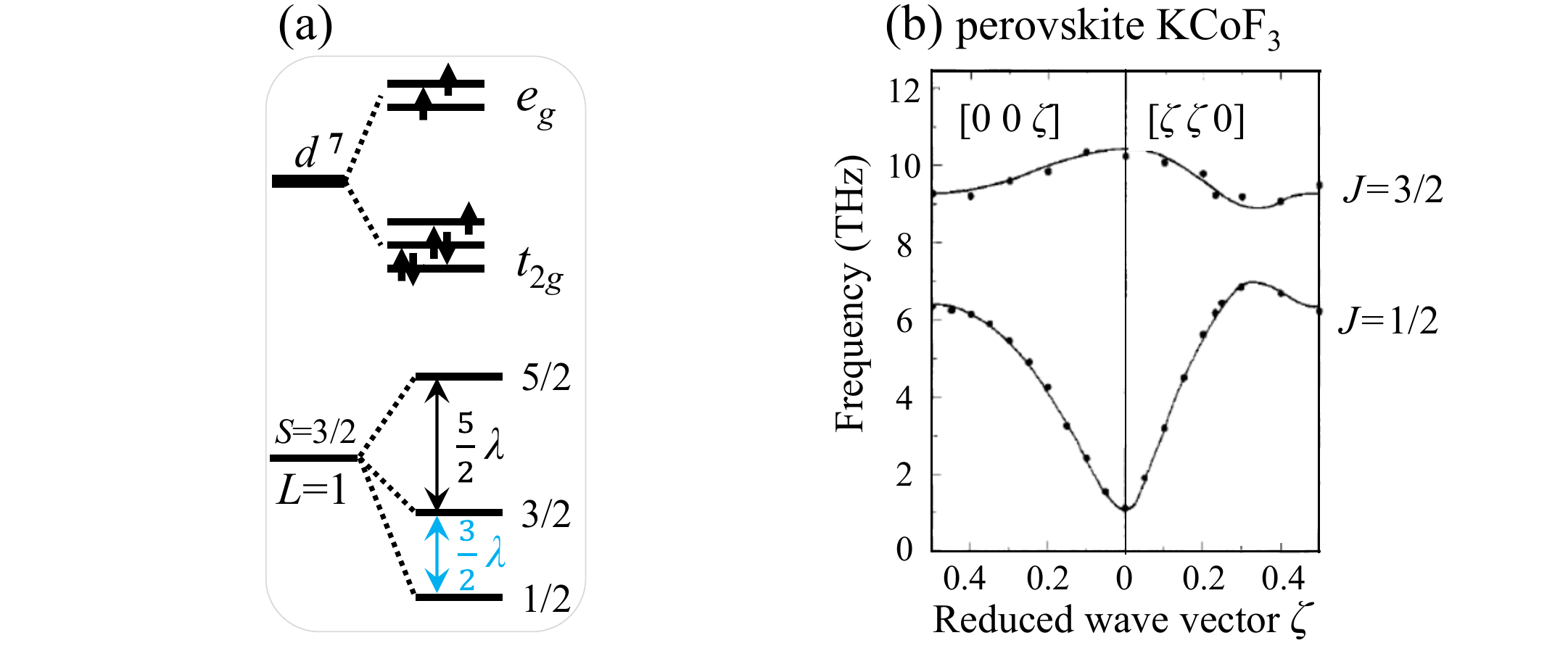}
\caption[]{(a) Level structures of the $d^7$ electronic configuration in an octahedral
crystal field without and with SOC. (b) Inelastic neutron  scattering measured
dispersion curves of the lowest two excitations in KCoF$_3$, figure is reproduced from Ref.~\refcite{Buy71}. }
\label{fig:3}
\end{center}
\end{figure}

In fact, $3d$-cobalt compounds such as CoO, KCoF$_3$, CoCl$_2$, etc. have been known as
canonical examples of the pseudospin-1/2 magnetism for decades\cite{Hol71,Buy71,Yam74,Gof95}.
In cobaltates, the $d^7$ ions Co$^{2+}$ in an octahedral crystal field have a predominantly $t_{2g}^5e_g^2$
configuration\cite{Yam74,Pra59} and form a high spin $S=3/2$ state.
The orbital degeneracy is three-fold and can be described by an effective $L=1$ moments\cite{Abr70}.
The $S=3/2$, $L=1$ configuration is split by SOC $\lambda \vc L \cdot \vc S$ with the states labeled according to the
total angular momentum $J=$1/2, 3/2 and 5/2, as shown in Fig.~\ref{fig:3}(a).
The ground state Kramers doublet again hosts the pseudospin $\widetilde{S}=1/2$ state,
which is similar to the case of $d^5$ ions Ru$^{3+}$ and Ir$^{4+}$ with $t_{2g}^5$($S=1/2$, $L=1$) configuration.
This guarantees the presence of the Kitaev exchange interaction on the symmetry grounds.

To have well defined pseudospin-1/2 ground state, $\lambda$ should be strong enough
to overcome the exchange interactions and non-cubic crystal fields. This might seem to be a problem for $3d$ materials where, unlike the cases of $4d$ or $5d$ ions, the
spin-orbit coupling strength is small.
The actual value of $\lambda$  can be quantified experimentally from
the transition between spin-orbit levels $1/2 \rightarrow 3/2$, termed as ``spin-orbit exciton'', with the energy difference $\sim \frac{3}{2}\lambda$
(the excitation energy can be affected by the crystal field which will be discussed later).

In perovskite KCoF$_3$, the spin-orbit exciton mode was observed
at $\sim 40$~meV by the inelastic neutron scattering~\cite{Hol71,Buy71}, see Fig.~\ref{fig:3} (b),
which is well separated from the low energy pseudospin-1/2 magnons.
In the quasi-two dimensional honeycomb lattices with less nearest neighbors, the magnon dispersion is expected to be narrower compared with
 perovskite lattices, as indeed observed in honeycomb cobaltates,
with the spin-orbit exciton modes located well above the pseudospin-1/2 magnons\cite{Yua20,Ell20,Son20}. 
Hence, the notion of ``pseudospin'' itself is physically well justified and the corresponding spin-orbit excitations are assumed
to have only perturbative effects on magnetic orders and fluctuations.

\section{Exchange Hamiltonian between pseudospin-1/2 in $3d^7$ systems}
\label{Sec:5}
Since the pseudospin-1/2 picture in $d^7$ cobaltates is verified, the low energy exchange Hamiltonian
between pseudospins should be of the same form as in $d^5$ systems. To obtain the corresponding
exchange constants $K$, $J$, $\Gamma$ and $\Gamma'$ in cobaltates, one has to:\\
1) derive first the exchange interactions operating in the full spin-orbital Hilbert space including both $t_{2g}$ and $e_g$ orbitals;\\
2) project these interactions onto the low energy pseudospin-1/2 sector.\\

On symmetry grounds, the pseudospin $\widetilde{S}=1/2$ exchange Hamiltonian in both
$d^7$ and $d^5$ systems should share the identical form as in Eq.~\ref{eq:H}.
However, the presence of additional, spin-active $e_g$ electrons in $d^7$ cobaltates is expected
to have a strong impact on the actual values of exchange parameters. In
particular, they should affect the strength of the Kitaev-type couplings
relative to other terms in the Hamiltonian\cite{Liu18,San18,Liu20}.
In this section, we will present the quantitative results of  detailed calculations of exchange constants in
$3d^7$ systems, and discuss
 the most important physical parameters tuning
the exchange Hamiltonian and ground states in honeycomb cobaltates.

\subsection{Role of $e_g$ electrons, charge-transfer vs Mott insulators}
\begin{figure}
\begin{center}
\includegraphics[width=12.6cm]{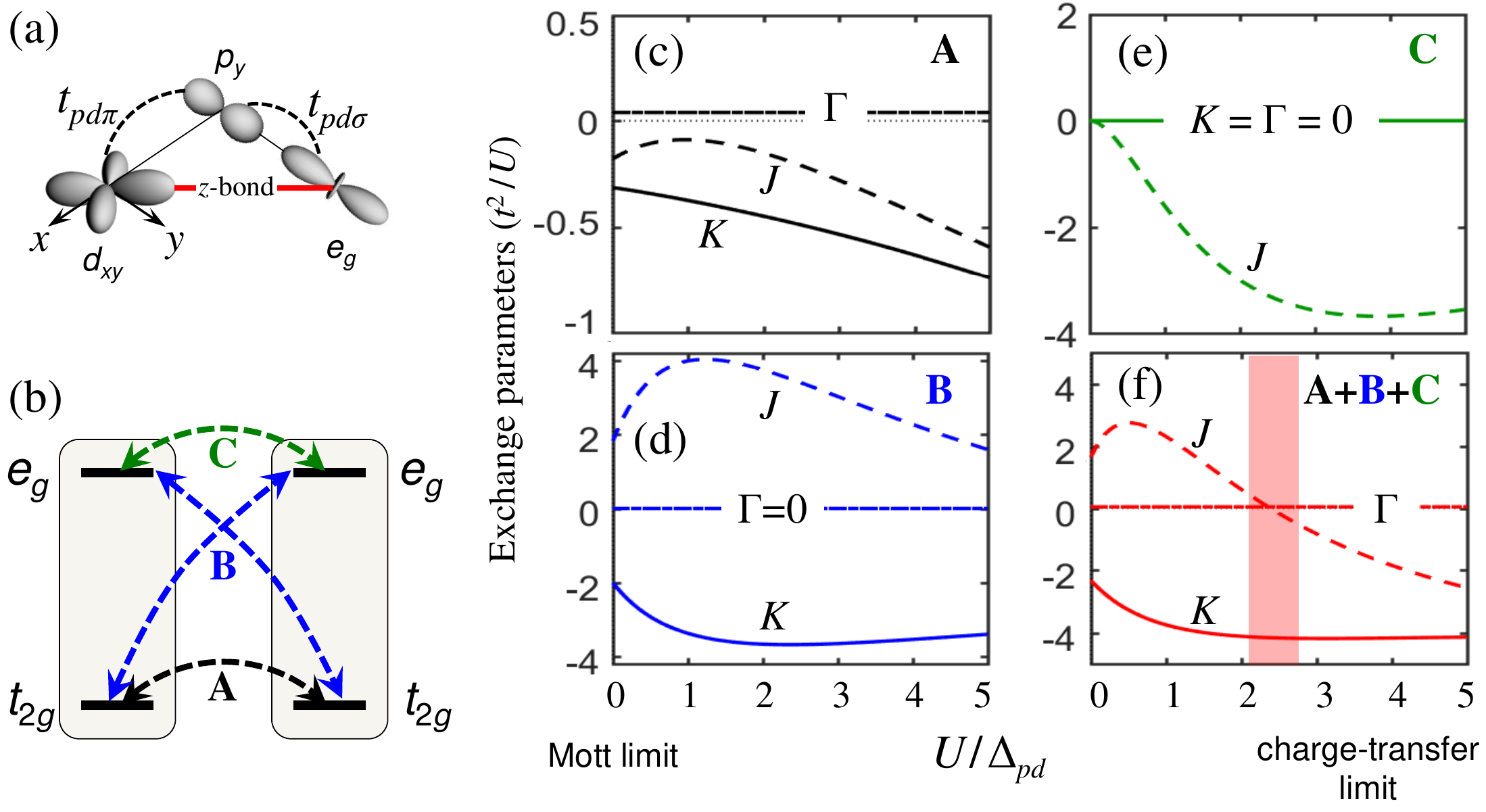}
\caption[]{(a) Sketch of $\pi$ and $\sigma$ hoppings involving $t_{2g}$ and $e_g$
orbitals along $z$-type NN bond. Only $c=d_{xy}$ orbital is active here.
(b) Three different classes of the exchange processes, A, B, and C.
(c-f) Exchange couplings $K$ (solid), $J$ (dashed) and $\Gamma$ (dash dotted) in units of $t^2/U$
as a function of $U/\Delta_{pd}$ contributed by hopping processes A (c), B (d), C (e), and total contribution (f).
$U$ is Coulomb interaction for $d$ ions and $\Delta_{pd}$
is $pd$ charge-transfer gap, and figure is reproduced from Ref.~\refcite{Liu18} with
$J_H/U=0.15$ and $t'/t=0.2$. The shaded area in (f) indicates the parameter space where $|K|
/\sqrt{J^2+\Gamma^2} \geq 8$.}
\label{fig:4}
\end{center}
\end{figure}

In the honeycomb 90$^{\circ}$ hopping geometry, the hopping integral associated with $e_g$ orbitals
is quite large since it involves the $\sigma$-type
hopping process $t_{pd\sigma}$($\sim 2t_{pd\pi}$), as shown in Fig.~\ref{fig:4}(a).
Therefore, it is essential to include the exchange processes related to $e_g$ orbitals.
Schematically, the exchange processes can be divided into three classes, as shown in Fig.~\ref{fig:4} (b),
by the exchange between \\
  A: $ \qquad $ $t_{2g}$ and $t_{2g}$ orbitals, \\
  B: $ \qquad $ $t_{2g}$ and $e_{g}$ orbitals, and \\
  C: $ \qquad $ $e_{g}$ and $e_{g}$ orbitals.

Identical to the $d^5$ systems, the hopping process A yields the extended
 $K$-$J$-$\Gamma$ model with the coupling constants shown in Fig.~\ref{fig:4} (c).
The $\Gamma$ term is expected to be rather weak
 due to the small $t'$ hopping in $3d$ system.
Regarding $J$ and $K$ couplings, they both remain FM due to Hund's coupling and their strength
 strongly depends on whether the system is in Mott ($U<\Delta_{pd}$) or charge-transfer ($U>\Delta_{pd}$) insulating
regime\cite{Zaa85}.

The process B involving spin active $e_g$ orbitals produces   $K$-$J$ model
with AFM $J>0$ and FM $K<0$, see Fig.~\ref{fig:4} (d).
The overall magnitude of the exchange couplings is stronger than that of process A
because of two reasons: the first one is due to the strong $\sigma$ hopping between $p$ and $e_g$ orbitals
as mentioned above; the second reason is that the pure $t_{2g}$ exchange couplings in process A originate from higher order contributions
and thus are small.
As a result, the final properties of the exchange Hamiltonian between pseudospin-1/2
for $d^7$ ions are given mostly by the contributions involving $e_g$ electrons.

Regarding the hopping between $e_g$ orbitals, process C gives pure Heisenberg model with FM $J<0$ as shown in Fig.~\ref{fig:4} (e).
This is expected from Goodenough-Kanamori rules\cite{Goo63} for orbitals that do not directly
overlap and interact via Hund's coupling on $p$ orbitals\cite{Liu18}. As a result, we have
\begin{align}
\left\{
\begin{aligned}
&A: \ \ \ t_{2g}{\text -} t_{2g}~(\rm weak) &\ \ \ & J<0,~K<0,~\Gamma \sim0; \\
&B: \ \ \ t_{2g}{\text -} e_g~(\rm strong)&        & J>0,~K<0,~\Gamma=0; \\
&C: \ \ \ e_g{\text -} e_g~(\rm strong)&           & J<0,~K=0,~\Gamma=0.
\end{aligned}
\right.
\end{align}
The contribution from channel C can compensate the strong AFM $J>0$ from channel B; this results in the dominance of strong FM Kitaev interaction $K$  in parameter region of $U/\Delta_{pd} \sim 2$-$3$, see Fig.~\ref{fig:4} (f).

Regarding $U/\Delta_{pd}$ ratio in cobalt compounds, this may vary broadly
depending on material chemistry, in particular on the electronegativity of
the anions. From the $ab$ $initio$ calculations, it is estimated $U\sim 5.0$-$7.8$~eV\cite{Ani91,Pic98,Jia10}.
While $\Delta_{pd} \sim 4$~eV in oxides, this value is much reduced in
compounds with Cl, S, P, etc.\cite{Zaa85,Bre86}, so that
$\Delta_{pd} \sim 2$-$4$~eV and $U/\Delta_{pd} \sim 2$-$3$ values seem to be plausible
in cobaltates.
The charge-transfer type cobalt insulators may indeed realize the situation when
Kitaev interaction dominates over isotropic Heisenberg coupling and $\Gamma$ term.

\subsection{Trigonal crystal field effect}
Commonly, the Jahn-Teller (JT) effect (``orbital-lattice coupling'') in pseudospin-1/2 systems is believed not essential at all,
since it cannot split the ground Kramers doublet.
However, the JT coupling breaks the symmetry and modifies the spatial shape of
the pseudospin wave functions through spin-orbit coupling.
By virtue of the pseudo-JT effect\cite{Ber06,Ber13}, the orbital-lattice coupling can be
converted into the pseudospin-lattice coupling.
The JT physics, even though rarely explored, is universal in spin-orbit Mott insulators and plays an important part
when describing the low energy physics such as in Sr$_2$IrO$_4$ and Ca$_2$RuO$_4$\cite{Liu19}.

Through the pseudospin-lattice coupling, the JT effect generates new terms or
renormalizes the coupling constants in the Hamiltonian\cite{Win17,Liu19,Bis19} through modifying the wave functions.
For instance, the symmetry allowed $\Gamma'$ term is originated from the
non-zero trigonal crystal field as shown in Eq.~\ref{eq:H}, suggesting
the crystal field as an efficient tuning parameter of the exchange couplings.
However, there is a concern that  the non-cubic crystal fields present in real materials may quench orbital moments and suppress the bond-dependence of the exchange couplings\cite{Kha05}.

\begin{figure}
\begin{center}
\includegraphics[width=12.6cm]{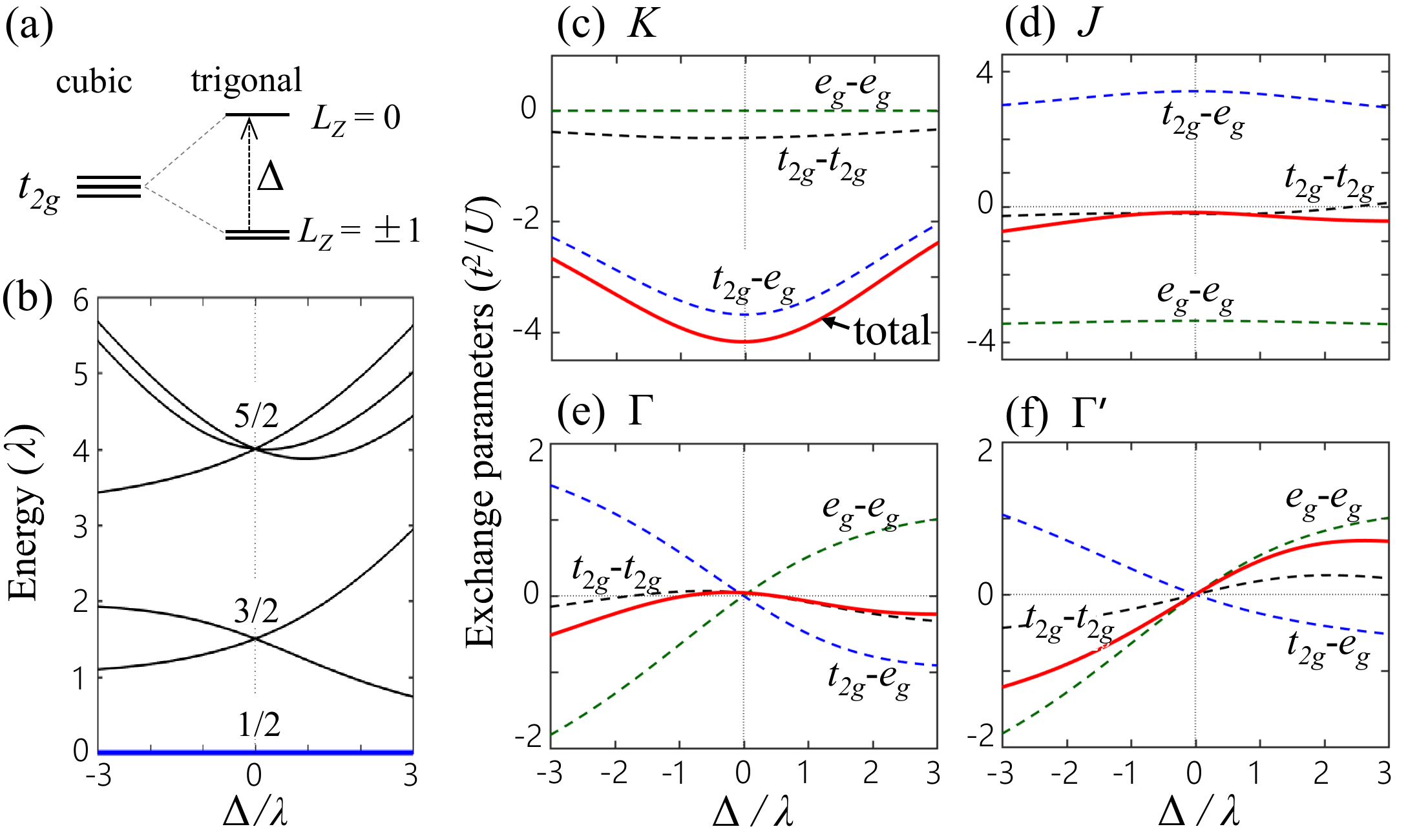}
\caption[]{(a) Splitting of $t_{2g}$-electron level under trigonal crystal field $\Delta$.
Within a point-charge model, $\Delta> 0$
 corresponds
to an elongation of octahedra along the trigonal $Z$-axis in Fig.~\ref{fig:1}.
(b) Splitting of $S=3/2$, $L=1$ manifold under spin-orbit coupling $\lambda$ and
trigonal field $\Delta$.
(c)-(f) Exchange parameters $K$, $J$, $\Gamma$, and $\Gamma'$ (red solid lines) as a function of $\Delta/\lambda$
and their individual contribution from $t_{2g}$-$t_{2g}$ (black), $t_{2g}$-$e_g$ (blue), and $e_g$-$e_g$ (green) exchange channels
with $U/\Delta_{pd}=2.5$, Hund's coupling $J_H=0.15U$ and $t'/t=0.2$.
The figure is taken from Ref.~\refcite{Liu20}.}
\label{fig:5}
\end{center}
\end{figure}

To address this issue, the trigonal crystal field effect in $d^7$ ions has been studied in Ref.~\refcite{Liu20}.
Since the trigonal distortion is defined in the $X$, $Y$ and $Z$ global coordinates
as shown in Fig.~\ref{fig:1}(a), it is easier to adopt the global coordinates as
the quantization axes. For instance, the effective angular momentum
$L_Z$-projections read as:
\begin{align}
|L_Z=0\rangle &=\frac{1}{\sqrt{3}} \left(|a\rangle+|b\rangle+|c\rangle \right),
\notag\\
|L_Z= \pm 1\rangle &= \pm \frac{1}{\sqrt{3}} \left(e^{ \pm i\tfrac{2\pi}{3}}|a\rangle
+e^{\mp i\tfrac{2\pi}{3}}|b\rangle+|c\rangle \right).
\end{align}
we recall that  $a=d_{yz}$, $b=d_{zx}$, and $c=d_{xy}$.
Following the steps explained above, we start with the wave function of the ground state.
Under trigonal crystal field $\Delta$, the three-fold $t_{2g}$
orbitals are split into one singlet and one doublet as shown in Fig.~\ref{fig:5}(a).
The pseudospin-1/2 ground state still preserves Kramers degeneracy, and its wave functions
written in the basis of $|S_Z,L_Z \rangle$ are:
\begin{align}
\Big| \!\widetilde{\frac{1}{2}},\pm \widetilde{\frac{1}{2}} \Big\rangle =
\mathcal{C}_1 \Big| \pm\frac{3}{2}, \mp 1 \Big\rangle + \mathcal{C}_2 \Big|
\pm \frac{1}{2},0 \Big\rangle+\mathcal{C}_3\Big| \mp \frac{1}{2}, \pm 1 \Big\rangle.
\label{eq:wf}
\end{align}
The coefficients obey a relation $\mathcal{C}_1:\mathcal{C}_2:\mathcal{C}_3 =
\frac{\sqrt{6}}{r_1}:-1:\frac{\sqrt{8}}{r_1+2}$, where the parameter $r_1>0$  is determined by the equation $\frac{\Delta}{\lambda}
=\frac{r_1+3}{2}-\frac{3}{r_1}-\frac{4}{r_1+2}$\cite{Lin63}.
At cubic limit, we have $(\mathcal{C}_1,\mathcal{C}_2,\mathcal{C}_3)
=(\tfrac{1}{\sqrt{2}},\tfrac{-1}{\sqrt{3}},\tfrac{1}{\sqrt{6}})$ which indicates
equal  contributions from $a=d_{yz}$, $b=d_{zx}$, and $c=d_{xy}$ orbitals.

As shown in Fig.~\ref{fig:5}(b), the $3/2$ quartet is split into two doublets with one doublet leans towards the
pseudospin-1/2 ground state. From the experiments in layered cobaltates, we know that the pseudospin $\widetilde{S}=1/2$ magnons ($\sim 10$~meV\cite{Yua20,Ell20,Yua20b,Son20,Che20,Lin20,Kim20}) are well separated from
higher lying spin-orbit excitations ($\sim 30$~meV\cite{Yua20,Ell20,Son20}),
indicating that the orbital moment is not yet  quenched and   the low energy physics can indeed be described using the pseudospin-1/2 language.

Then, the NN exchange Hamiltonian can be derived using the wave function Eq.~\ref{eq:wf}
under the trigonal crystal field\cite{Liu20}. Since $\Delta$ does not break the in-plane $C_3$ symmetry,
the  obtained exchange Hamiltonian is of the same form as in Eq.~\ref{eq:H} but with renormalized parameters.
Similar with the cubic case, Kitaev coupling $K$ comes almost entirely from the $t_{2g}$-$e_g$ process,
and is still dominant within finite $\Delta$ regime as long as the pseudospin-1/2 picture stays valid, see Fig.~\ref{fig:5}(c) and also discussion in Sec.~\ref{Sec:6} below.
Acting via modification of the pseudospin wave function Eq.~\ref{eq:wf}, the trigonal field $\Delta$
has especially strong impact  on the non-Kitaev couplings $J$, $\Gamma$, $\Gamma'$, as shown in Figs.~\ref{fig:5}(d)-\ref{fig:5}(f).
This suggests that $\Delta$ could be served as an efficient and also experimentally accessible parameter
that controls the  relative strength of these ``undesired'' terms.
It is also noticed that $t_{2g}$-$e_g$ and $e_g$-$e_g$ contributions to $J$, $\Gamma$, and $\Gamma'$ are of
opposite signs and largely cancel each other, resulting in overall small values of these couplings.

\subsection{Combined effects of $U/\Delta_{pd}$ and $\Delta/\lambda$ on exchange parameters}
\begin{figure}
\begin{center}
\includegraphics[width=12.6cm]{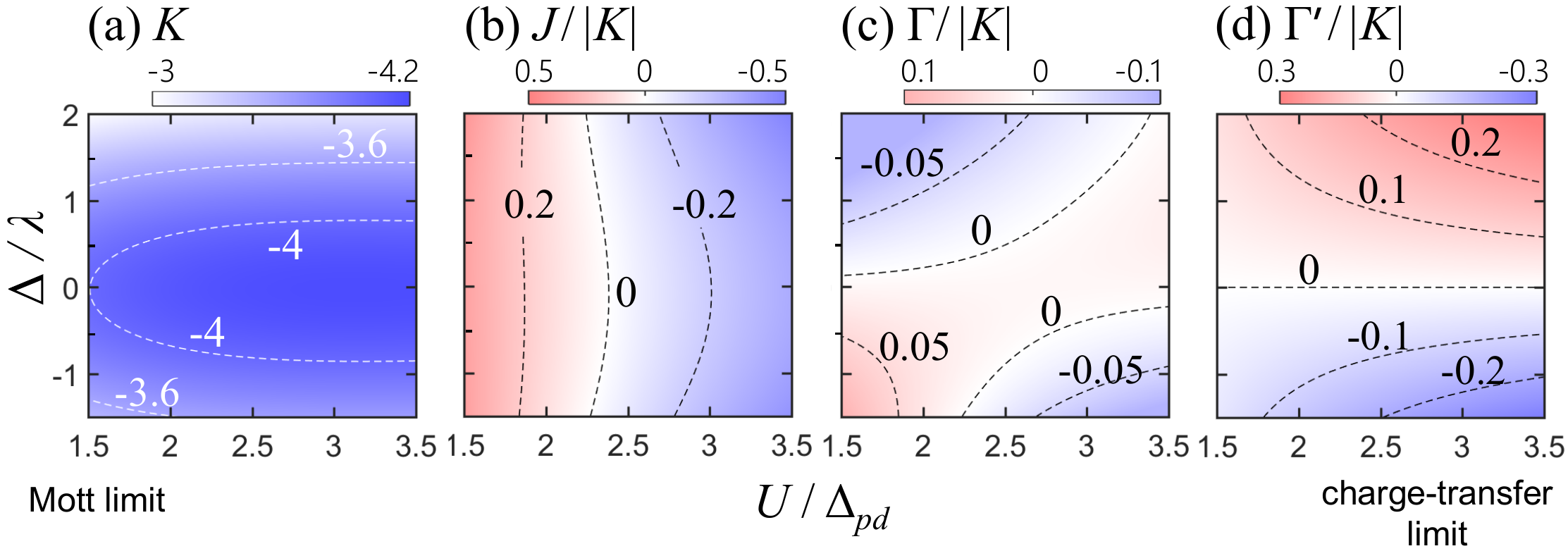}
\caption[]{(a) Kitaev coupling $K$ (in units of $t^2/U$), and (b)-(d) the relative values of
$J/|K|$, $\Gamma/|K|$, and $\Gamma'/|K|$ as a function of $\Delta/\lambda$ and $U/\Delta_{pd}$.
Figure is taken from Ref.~\refcite{Liu20}.}
\label{fig:6}
\end{center}
\end{figure}

From the above discussions, it is clear that $U/\Delta_{pd}$ and $\Delta/\lambda$ are two important physical
parameters that decide the values of the NN exchange constants. We present in Fig.~\ref{fig:6}
the combined effect  of these two parameters on the exchange constants.

Within the parameter  space shown in Fig.~\ref{fig:6} (a), FM Kitaev coupling $K<0$ is dominant.
The Kitaev coupling $K$ is not much sensitive to either $U/\Delta_{pd}$ or $\Delta/\lambda$ variations,
providing a robust foundation of realizing the Kitaev physics in $3d$ cobaltates.
On the other hand, the Heisenberg coupling $J$ is more sensitive to $U/\Delta_{pd}$ rather than $\Delta/\lambda$, see Fig.~\ref{fig:6}(b).
$\Gamma$ term in Fig.~\ref{fig:6}(c) is the weakest interaction due to the small direct hopping $t'$ between $3d$ orbitals
(as compared to the extended $4d$ or $5d$ orbitals).
The trigonal crystal field generated $\Gamma'$ interaction is not much sensitive to $U/\Delta_{pd}$ ratio, see Fig.~\ref{fig:6}(d).

\subsection{Magnetic phase diagram}
Having quantified the exchange parameters in Hamiltonian Eq.~\ref{eq:H}, the corresponding ground states
can be addressed in $d^7$ systems. The obtained model is highly frustrated since the leading term is the bond-dependent
Kitaev coupling. Therefore, the ED method\cite{Cha10,Cha13,Rau14,Oka13,Cha16,Rus19} is employed to
study the phase behavior under $U/\Delta_{pd}$ and $\Delta/\lambda$.

\begin{figure}
\begin{center}
\includegraphics[width=12.6cm]{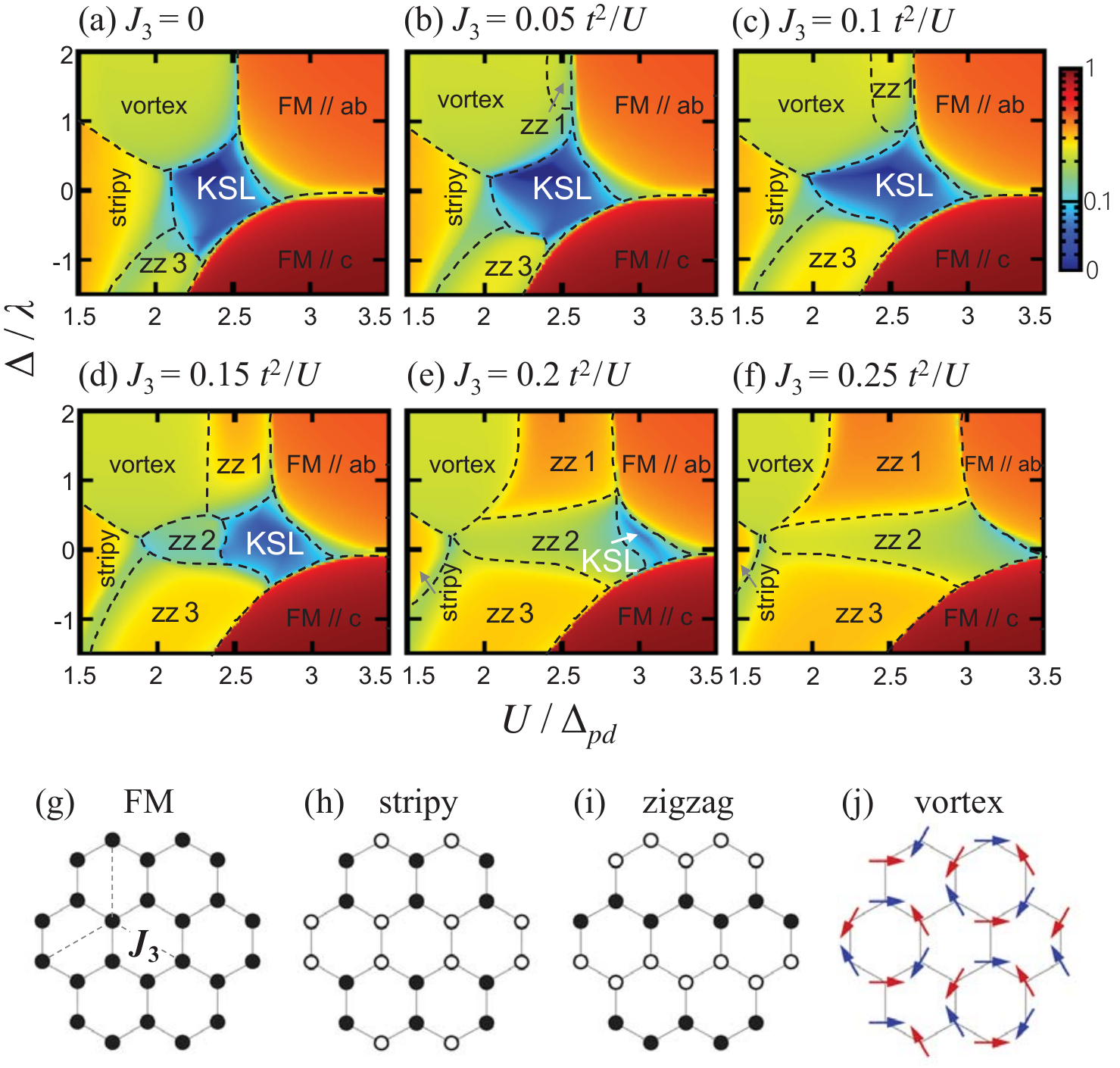}
\caption[]{The phase diagram obtained by ED on a hexagon-shaped 24-site cluster
with  the third-
NN Heisenberg interaction $J_3$ values equal to (a) 0, (b) $0.05$, (c) $0.1$, (d) $0.15$, (e) $0.2$ and (b) $0.25$ in units of $t^2/U$.
The zigzag-type states have three different types with moments in the $ab$ plane (zz1), along one of the three local axes (zz2), and in the $ac$ plane (zz3).
The color map shows the second-NN spin correlation strength which drops sharply in the KSL phase.
Sketch of the magnetic structures for (e) FM, (f) zigzag, (g) stripy, and (h) vortex orders. Open and closed circles represent opposite spin directions.
Figures are taken from Ref.~\refcite{Liu20}.}
\label{fig:7}
\end{center}
\end{figure}

As shown in Fig.~\ref{fig:7} (a), the KSL phase is stabilized at the center of the parameter space, where the non-Kitaev couplings are small (roughly $< 10\% |K|$).
Surrounding the KSL phase, there are two types of FM orders, stripy, zigzag and finally a vortex phase as depicted in Figs.~\ref{fig:7}(g)-\ref{fig:7}(j). The two FM orders  differ by the alignment of the magnetic moments, it is in the honeycomb plane for FM//ab, while it is perpendicular to the honeycomb plane for FM//c. The phase boundary for two FM orders is approximately at $\Delta=0$, in other words, the sign of $\Gamma'$ decides the moment directions of FM orders.
For the zigzag phase labeled ``zz3", the magnetic moments are in the $XZ$ (ac) plane as in Na$_2$IrO$_3$\cite{Chu15,Cha16} and $\alpha$-RuCl$_3$\cite{Cao16,Sea20}.

At cubic limit ($\Delta=0$) where $\Gamma'=0$ exactly,
the obtained Hamiltonian can be approximated as the well studied $K$-$J$ model once the rather weak $\Gamma \sim 0$ term is ignored.
While changing the $U/\Delta_{pd}$ ratio, $J$ changes from AFM $J>0$ to FM $J<0$
and the Kitaev interaction $K$ remains FM.
Consequently, the ground state changes from stripy to FM order through the intermediate KSL phase\cite{Cha13} as in
Fig.~\ref{fig:2}(a) and Fig.~\ref{fig:2}(b).

When the trigonal field $\Delta$ is switched on, the $\Gamma'$ term is activated and
confines the KSL phase to the window of $|\Delta|/\lambda<1$ where $|\Gamma'/K|<0.1$.
When it is close to the Mott limit where Heisenberg coupling $J>0$,
the stripy state gives way to a vortex-type magnetic order at positive $\Delta$,
and to the zigzag order for negative $\Delta$ due to the combined effect of $\Gamma$ and $\Gamma'$ terms.
At the charge-transfer limit, there are two types of FM orders with the direction of the magnetic moments
decided by the sign of $\Gamma'$.

In summary, the NN pseudospin-1/2 exchange Hamiltonian in $d^7$ systems is dominated by the FM Kitaev model,
which is robust against the trigonal splitting of orbitals (see also the detailed discussion in  Sec.~\ref{Sec:6} below).
The ``non-Kitaev'' terms represented mostly by $J$ and $\Gamma'$ couplings shape
the phase diagram, and constrain the KSL phase within the area when the Kitaev term is roughly 10 times
larger than the other couplings.

\subsection{Role of third-NN Heisenberg coupling $J_3$}
The longer range exchange interactions, which are normally of the Heisenberg type, are unavoidable in real materials.
It is crucial to inspect how the above picture is modified by longer range interactions,
especially by the third-NN Heisenberg coupling $J_3\widetilde{\vc S}_i\cdot \widetilde{\vc S}_j$,
which appears to be one of the major obstacles on the way to a KSL in 5$d$ and 4$d$ compounds\cite{Win17,Win16}.
It is difficult to estimate the precise value of $J_3$ analytically, since long-range interactions involve
multiple exchange channels and are thus sensitive to material chemistry details. Hence they have to be determined
experimentally or through comprehensive density functional theory (DFT) calculations. Given the fact that there are very few experimental data quantifying the magnitudes of $J_3$
in honeycomb cobaltates, various values of $J_3$ are added ``by hand'' in the ED calculations.

The ground states have been re-examined with different $J_3$ values in Ref.~\refcite{Liu20} and
the modified phase diagram are shown in Figs.~\ref{fig:7}(b)-(f).
Compared with Fig.~\ref{fig:7}(a), two new zigzag phases ``zz1" and ``zz2" around the KSL phase are successively formed by the increased $J_3$.
This can be understood by considering the correlations of third NN in the zigzag phase,
which is characterized by AF oriented spins on all third-neighbor bonds, see Fig.~\ref{fig:7}(i).
Similarly, a large suppression may be
expected for FM and stripy phases that have FM aligned third NN spins.
The effect on the vortex phase is weak as each spin has one FM aligned third NN  and two
third NNs at an angle of $120^\circ$, leading to a
cancelation of $J_3$ in energy on the classical level.

In the large area covered by the zigzag order, various ratios and combinations
of signs of the NN interactions are realized. This is the origin
of three distinct zigzag phases zz1, zz2, and zz3, differing by their moment directions.
Negative $\Gamma$ and positive $\Gamma'$ found in zz1 phase space
lead to the in-plane moment direction.
The zz3 phase is characterized by positive $\Gamma$ and negative $\Gamma'$
interactions which stabilizes the zigzag order, as in the case of Na$_2$IrO$_3$\cite{Chu15,Cha16,KimCha20}.
Finally, in the zz2 phase, $\Gamma$ and $\Gamma'$ terms maintain only small
values and moment directions pointing along cubic axes $x$, $y$ or $z$, which is selected by order-from-disorder mechanism\cite{Cha16}.

In the KSL phase where the third NN spins are not correlated at all,
small $J_3$ has a moderate negative impact when trying to align
them in AF fashion. After including nonzero $J_3$, the KSL phase slightly grows first at the expense of FM and stripy phases, as shown in Figs.~\ref{fig:7}(b,c).
At the same time, the KSL phase is  expelled from the bottom left corner by the expanding of zz3 phase to the right
where FM $J$ and AFM $J_3$ tend to frustrate each other.
Once $J_3$ reaches $0.25t^2/U$ ($|J_3/K|\sim 0.06$) \cite{Liu20},
the zigzag order quickly takes over, suppressing the KSL phase completely in Fig.~\ref{fig:7}(f).

It is clear that $J_3$ plays a very important role in determining the magnetic properties,
and the existence of KSL phase is very sensitive to $|J_3/K|$ ratio.
We note that $|J_3/K|\simeq 0.1$ was estimated\cite{Win18,Win17a,Suz20} in the 4$d$ compound $\alpha$-RuCl$_3$.
In principle, this ratio is expected to be smaller in cobaltates with more localized 3$d$ orbitals,
cf.  the radial extension of the wavefunctions $\langle r^2 \rangle_{3d}=1.25$
for Co$^{2+}$ and $\langle r^2 \rangle_{4d}=2.31$ for Ru$^{3+}$ ions (in atomic units),
respectively\cite{Abr70}. The estimated hopping integral for third NN is 10\% of that for first NN in
honeycomb CoTiO$_3$\cite{Ell20}. Thus, in principle it is promising that the $|J_3/K|$ ratio in cobaltates can be below the critical
value of eliminating the existence of KSL phase.
Yet, the magnitudes of $J_3$ in honeycomb $d^7$ materials still need to be quantified experimentally or via systematic DFT calculations
similar with what has been done for $5d$ Na$_2$IrO$_3$\cite{Foy13}.

\subsection{Robustness of the   Kitaev interactions in cobaltates}
\label{Sec:6}
\begin{figure}
\begin{center}
\includegraphics[width=12.6cm]{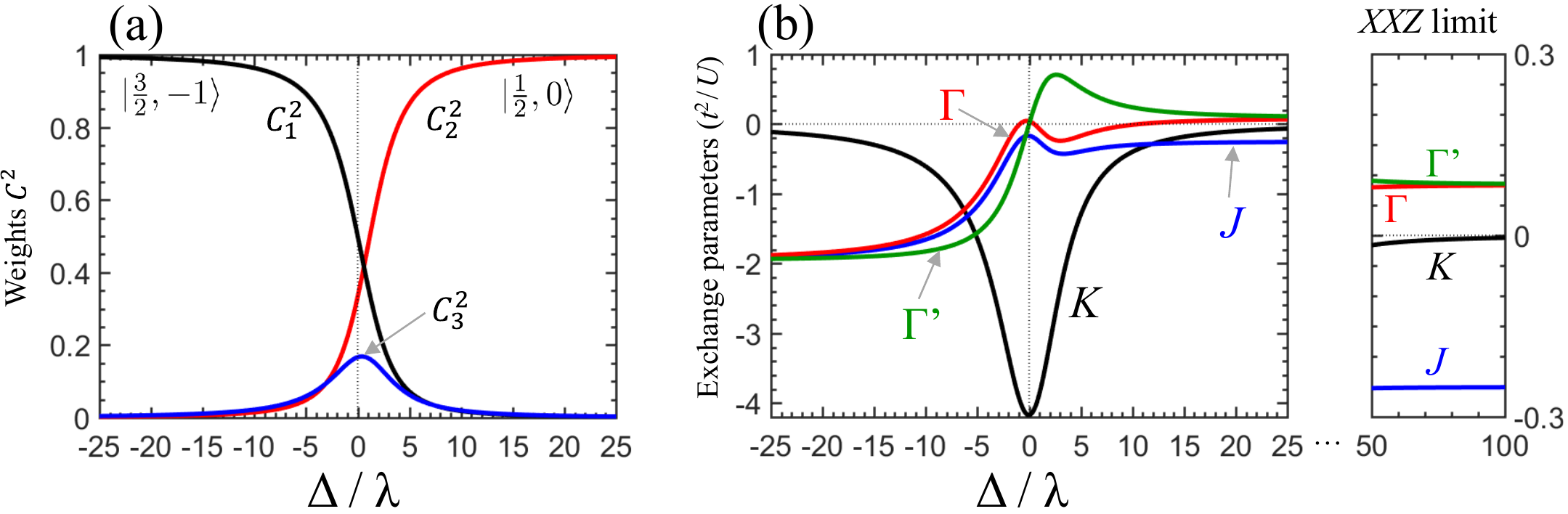}
\caption[]{(a) Weights $\mathcal{C}_1^2$ (black), $\mathcal{C}_2^2$ (red), and $\mathcal{C}_3^2$ (blue)
of different components $|S_Z,L_Z\rangle$ in the ground state pseudospin-1/2 wave function Eq.~\ref{eq:wf}.
Only at very large values of $\Delta/\lambda$, when either $\mathcal{C}_1 \simeq 1$ or $\mathcal{C}_2 \simeq 1$,
a single component product state $|S_Z,L_Z\rangle$ can be realized.
(b) Exchange parameters $K$ (black), $J$ (blue), $\Gamma$ (red) and $\Gamma'$ (green) as a function of $\Delta/\lambda$.
A conventional $XXZ$ model with $K=0$ and $\Gamma=\Gamma'$ is realized only at very large $\Delta/\lambda$, as illustrated on the right-hand side as an example.
}
\label{fig:8}
\end{center}
\end{figure}

It is found above that $K$ term is dominant in the range where $\Delta$ is comparable with $\lambda$.
However, it is important to examine how the Kitaev term evolves at very large crystal fields.
In other words, we would like to see the limitations of the Kitaev model description in cobaltates.
To this end, the calculations are extended to large trigonal field regime, and the results are shown in Fig.~\ref{fig:8}.

For $|\Delta/\lambda|<5$ roughly, the orbital moment is not fully quenched and the
wave functions of the ground state are coherent superpositions of spin-orbit entangled states, see Fig.~\ref{fig:8}(a).
When  $|\Delta/\lambda|$ is increased, the orbital degeneracy is lifted and the entanglement between spin and orbital is suppressed, then the pseudospin wavefunction becomes a single component product state.

The degree of spin-orbit entanglement in the ground state dictates the relative
strength of Kitaev coupling.
For $|\Delta/\lambda|<5$, where the spin and orbital are highly entangled,
$K$ coupling remains the largest among the others, as shown in the left panel of Fig.~\ref{fig:8} (b).
With further increased $|\Delta/\lambda|$, the non-Kitaev interactions become comparable with $K$ term.
At very large $|\Delta/\lambda| > 10$, one observes $K \sim 0$ and $\Gamma \simeq \Gamma'$.
In this limit, bond-directional nature is quenched by the crystal field, and the model becomes similar to a conventional $XXZ$ model which was
commonly adopted to analyze the experimental data in Co$^{2+}$ compounds\cite{Yua20,Hut73,Reg06,Tom11,Ros17,Nai18}.

To see the relation between $XXZ$ model and general Hamiltonian Eq.~\ref{eq:H}, it is helpful to rewrite the latter in hexagonal coordinate axes $XYZ$ frame\cite{Cha15}:
\begin{align}
\mathcal{H}_{ij}^{(\gamma)}= & J_{XY}
\left(\widetilde{S}_i^X\widetilde{S}_j^X +
\widetilde{S}_i^Y\widetilde{S}_j^Y \right) +
J_Z \widetilde{S}_i^Z\widetilde{S}_j^Z  \notag \\
+ & A \left[ c_{\gamma}\left(\widetilde{S}_i^X\widetilde{S}_j^X -
\widetilde{S}_i^Y\widetilde{S}_j^Y \right)-
s_{\gamma}\left(\widetilde{S}_i^X\widetilde{S}_j^Y +
\widetilde{S}_i^Y\widetilde{S}_j^X \right) \right]   \notag \\
- & B \sqrt {2}\left[ c_{\gamma}
\left(\widetilde{S}_i^X\widetilde{S}_j^Z +
\widetilde{S}_i^Z\widetilde{S}_j^X \right)
+s_{\gamma}\left(\widetilde{S}_i^Y\widetilde{S}_j^Z +
\widetilde{S}_i^Z\widetilde{S}_j^Y \right) \right] ,
\label{eq:HJ}
\end{align}
with
$c_{\gamma}\equiv \cos \phi_{\gamma}$ and $s_{\gamma}\equiv \sin \phi_{\gamma}$.
The angles $\phi_{\gamma}=0, \tfrac{2 \pi }{3}, \tfrac{4 \pi }{3}$ refer to the
$z$-, $x$-, and $y$-type bonds, respectively.
The transformations between the two sets of parameters
entering Eq.~\ref{eq:H} and Eq.~\ref{eq:HJ} are:
\begin{alignat}{2}
&J_{XY} =J+\tfrac{1}{3}K-\tfrac{1}{3}(\Gamma+2\Gamma') \; ,
&\quad \ \ \ \ \ \ \ \ \ \ \ \ \
&K =A+2B \; ,
\notag \\
&J_Z =J+\tfrac{1}{3}K+\tfrac{2}{3}(\Gamma+2\Gamma')\; , &
&J =\tfrac{1}{3}(2J_{XY}+J_Z-A-2B) \; ,
\notag \\
&A =\tfrac{1}{3}K+\tfrac{2}{3}(\Gamma-\Gamma') \; , &
&\Gamma =\tfrac{2}{3}(A-B)+\tfrac{1}{3}(J_Z-J_{XY})  \; ,
\notag \\
&B =\tfrac{1}{3}K-\tfrac{1}{3}(\Gamma-\Gamma')  \; , &
&\Gamma' =\tfrac{1}{3}(J_Z-J_{XY}+B-A) \; .
\label{eq:Tr}
\end{alignat}

The $XXZ$ model corresponds $A=B=0$ in Eq.~\ref{eq:HJ} indicating the Kitaev-type anisotropy
disappears (i.e. $K=0$) and also $\Gamma = \Gamma'$,
which will only be realized when $|\Delta/\lambda| \gg 10$, see the right panel of Fig.~\ref{fig:8}(b) for example.
However, such an extreme limit is unlikely for realistic trigonal fields.
Thus, a proper description of magnetism in cobaltates should be based on the model of Eq.~(\ref{eq:H})
accounting for the bond-directional nature of pseudospin-1/2 interactions. 

\subsection{$3d$ cobaltates versus $4d$ and $5d$ compounds}
\label{sec:5}

Before moving forward, a temporary summary could be made here.
A general form of the exchange Hamiltonian Eq.~\ref{eq:H} is established in cobaltates, which is the same as in $4d$ and $5d$ systems.
From a materials perspective, it is nice to extend the search area to cobaltates, especially given that Co is abundant and less expensive element compared with Ru and Ir.
More importantly, there are fundamental differences related to different electronic structures and exchange mechanisms, which make the proposal of realizing Kitaev physics in $3d$ systems very promising for the following reasons:

1) The presence of spin active $e_g$ electrons leads to strong reduction of non-Kitaev couplings, which results in the dominance of FM Kitaev term. Since the exchange between
 $e_g$ electrons is highly sensitive to the bond angle, this makes it possible to eliminate the destructive effects of NN non-Kitaev terms
 through lattice control, e.g., strain engineering.

2) The $3d$ orbitals are more localized than $4d$ and $5d$ ones, thus the ``unwanted'' long-range
interactions  $J_3$ and non diagonal $\Gamma$ terms 
should be smaller in $3d$ systems.

3) The decisive tuning parameter of the exchange Hamiltonian is the ratio of $\Delta/\lambda$.
Since the spin-orbit coupling strength is smaller in $3d$ ions
than in $4d$ and $5d$ ones, it is easier to manipulate the ground state wave function  and thus the exchange Hamiltonian parameters.

\section{$d^7$ Honeycomb Materials}
\label{sec:7}
So far, the presented theoretical results are encouraging. Now it is time to inspect the real materials.
Quite a number of such quasi-two-dimensional honeycomb magnets are known such as $A_3$Co$_2$SbO$_6$ ($A$=Na,Ag,Li)\cite{Son20,Kim20,Vic07,Won16,Yan19,Zve16,Str19,Viv20},
Na$_2$Co$_2$TeO$_6$\cite{Son20,Che20,Lin20,Kim20,Vic07,Lef16,Ber17,Yao20,Hon21},
BaCo$_2$($X$O$_4$)$_2$ ($X$=As, P)\cite{Reg06,Nai18,Zho18,Zho20},
CoTiO$_3$\cite{Yua20,Ell20,Yua20b,New64,Bal17}, CoPS$_3$\cite{Bre86,Wil17}, A$_2$Co$_4$O$_9$\cite{Ber61} (A=Nb, Ta)
and so on, see also the recent review Ref.~\refcite{Mot20b}.
Apart from honeycomb lattice compounds,
there are many $d^7$ cobaltates possessing pseudospin-1/2 ground state,
such as quasi-one dimensional CoNb$_2$O$_6$\cite{Col10,Mor21}, triangular lattice
antiferromagnets Ba$_3$CoSb$_2$O$_9$\cite{Zho12} and
Ba$_8$CoNb$_6$O$_{24}$\cite{Raw17}, spinel GeCo$_2$O$_4$\cite{Tom11}, and
pyrochlore lattice NaCaCo$_2$F$_7$\cite{Ros16,Ros17}. Several materials among the above have been
suggested to be proximate to the KSL phases.

In the layered honeycomb lattice structures, the inter-layer couplings, which involve rather long distance
and indirect exchange paths, are expected to be small. For instance, in Na$_2$Co$_2$TeO$_6$,
the inter-layer coupling is estimated about $1\%$ of the in-plane coupling from
magnetic Bragg peak lineshapes\cite{Lef16}.
This is similar with the honeycomb ruthenates such as $\alpha$-RuCl$_3$\cite{Kim16} and SrRu$_2$O$_6$\cite{Suz19}.
The small inter-layer couplings in Kitaev materials can indeed be neglected.
Its only role is to set up the c-axis coherence below long-range ordering temperature.
Thus, in the following discussions, we consider the two-dimensional model within the honeycomb plane.

The above listed honeycomb cobaltates are magnetically ordered at finite temperatures, with zigzag and FM orders being the most common phases
within the $ab$-plane which correspond to zz1 and FM $//$ ab phases discussed here.
If one can locate certain material in the phase diagram of Fig.~\ref{fig:7},
it will be clear  how to drive the magnetically
ordered state into the KSL phase by tuning an appropriate physical parameter.

To determine the exact position of a given material in the phase space of Fig.~\ref{fig:7}, three parameters are needed:
$\Delta/\lambda$, $U/\Delta_{pd}$ and $J_3$.
In this section, we will present how to estimate the three parameters and map the real materials onto the phase diagram.

\subsection{Quantifying the trigonal crystal field $\Delta$}

By measuring the excitation energy of the $1/2~\rightarrow~3/2 $ transition or the splitting of the excited multiplets,
the value of the crystal field $\Delta$ can be estimated together with the strength of SOC $\lambda$\cite{Yua20,Ell20}.
As an alternative option, $\Delta$ can be obtained from paramagnetic susceptibility $\chi^{\alpha}(T)$ ($\alpha=ab$ or $c$),
as we discuss in more detail now.

The free ion magnetic susceptibility per ion along certain direction $\alpha$ is:
\begin{align}
\label{eq:chi_ion}
\chi^{\alpha}_{\rm ion}=\frac{1}{Z(T)}\sum_{n,m}\frac{e^{-\beta E_n}-
e^{-\beta E_m}}{E_m-E_n} (M_{nm}^\alpha)^2 ,
\end{align}
the partition function $Z(T)= \sum_{n}e^{-\beta E_n}$, and
$\beta=\tfrac{1}{k_B T}$ where $k_B$ is the Boltzmann constant.
The integer numbers $n$ and $m$ run over all the 12 states in Fig.~\ref{fig:5}(b),
$M_{nm}^\alpha=\langle n|M_{\alpha}|m \rangle$ is matrix element of the magnetic moment operator
$\vc M= (2\vc S-\tfrac{3}{2}\kappa \vc L)$ (in units of Bohr magneton $\mu_B$) and $\kappa$ is the covalency reduction factor\cite{Abr70}.

\begin{figure}
\begin{center}
\includegraphics[width=12.6cm]{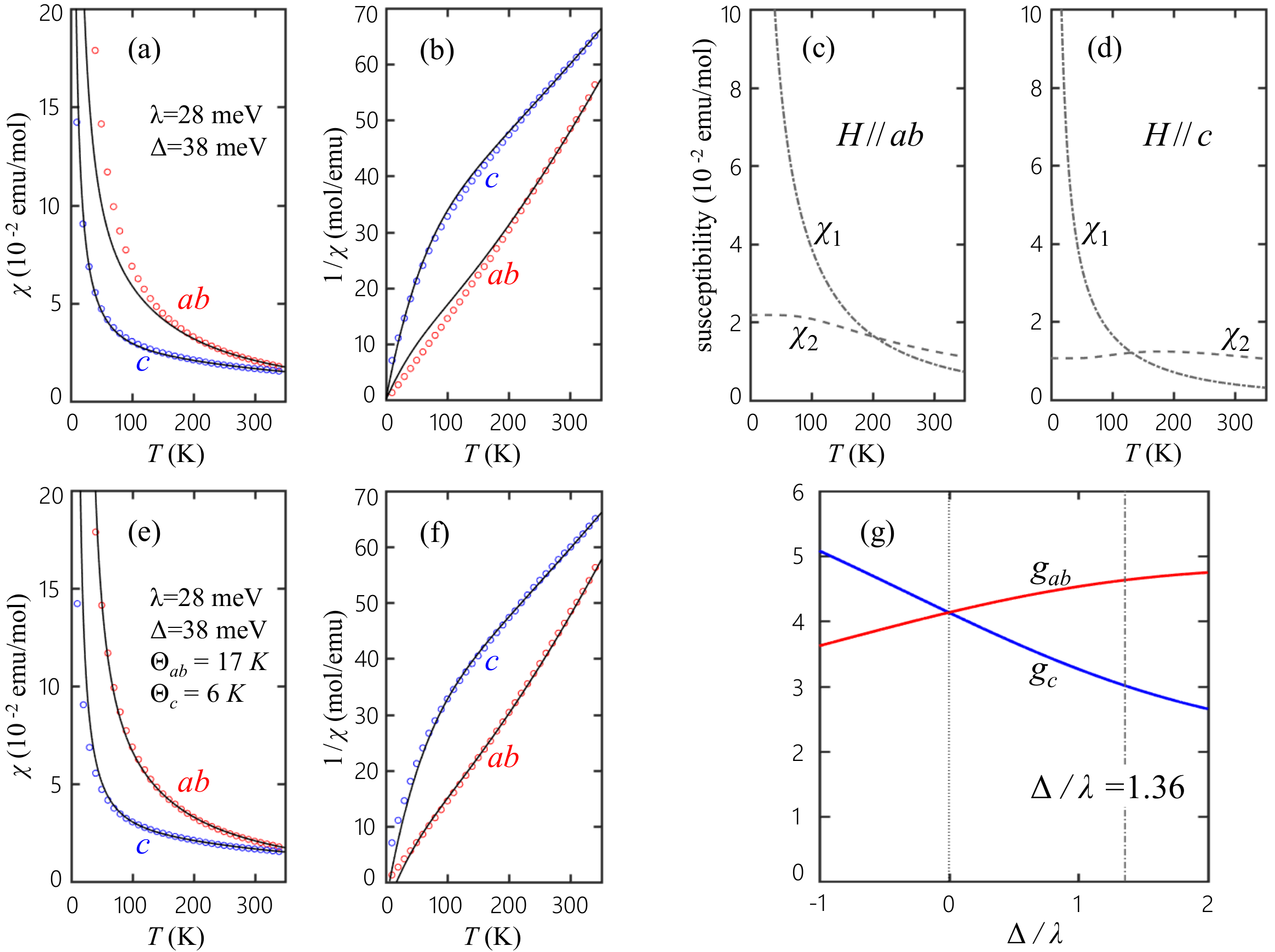}
\caption[]{(a)-(f) Temperature dependence of magnetic susceptibilities and its inverse in Na$_3$Co$_2$SbO$_6$.
(g) The g-factors $g_{ab}$ (red) and $g_c$ (blue) as a function of $\Delta/\lambda$. $\Delta/\lambda=1.36$ corresponds to Na$_3$Co$_2$SbO$_6$. Figure is taken from Ref.~\refcite{Liu20}.}
\label{fig:9}
\end{center}
\end{figure}

As an example, the experimental data of
Na$_3$Co$_2$SbO$_6$ from Ref.~\refcite{Yan19} has been fitted using $\chi^{\alpha}(T)=\chi^{\alpha}_{\rm ion}+\chi^{\alpha}_0$,
where $\chi^{\alpha}_0$ is a temperature independent constant.
Fair agreements with experiments for both $\chi^{ab}$ and $\chi^c$ can be obtained using $\Delta=38$~meV for
Na$_3$Co$_2$SbO$_6$ with $\lambda=28$~meV, see Figs.~\ref{fig:9}(a,b).

In Fig.~\ref{fig:9}(b), there is one important characteristic feature of the changes in the slopes of both
$1/\chi^{ab}$ and $1/\chi^c$ data.
To understand this, it is instructive to divide $\chi^{\alpha}_{\rm ion}$
into two parts, $\chi^{\alpha}_{\rm ion}=\chi^{\alpha}_1+\chi^{\alpha}_2$, where $\chi^{\alpha}_1$ term
accounts for the transitions within $\widetilde{S}=1/2$ doublet:
 \begin{align}
\label{eq:chi1}
\chi^{\alpha}_1= p_{1/2} \; \frac{(\mu^{\alpha}_{\rm eff})^2}{3k_B} \frac{1}{T} \; .
\end{align}
Here, $p_{1/2}=2/Z(T)$ measures the occupation of the ground state and the effective moments are given by $\mu^{\alpha}_{\rm eff}=g_{\alpha} \sqrt{\widetilde{S} (\widetilde{S}+1)}$. $\chi^{\alpha}_2$ is the Van-Vleck contribution of the excited states.
Since the excited levels of Co$^{2+}$ are relatively low, the weight $p_{1/2}$ of the Curie term as well as Van-Vleck contribution $\chi^{\alpha}_2$
are sensitive to the temperature $T$. In fact, this is exactly the reason why one should use the general form of the single-ion susceptibility Eq.~\ref{eq:chi_ion},
instead of the standard Curie-Weiss fit $\chi(T)=C/(T-\Theta)+\chi_0$ where the Curie constant $C$ is assumed to be temperature independent.
The characteristic changes in the slopes of $1/\chi^{ab}$ ($1/\chi^c$) around 200~K (100~K)
originate from the interplay between $\chi_1(T)$ and $\chi_2(T)$ which become of similar order at these temperatures, see Figs.~\ref{fig:9}(c,d).
In fact, this behavior is common also for other cobaltates (see Fig.~\ref{fig:13} and Fig.~\ref{fig:14} below).

There are apparent deviations of the fitting susceptibility at low temperatures in Figs.~\ref{fig:9}(a,b).
To solve this problem, one can include  correlations between the pseudospins in a molecular field approximation.
The Curie term $\chi^{\alpha}_1$ can be then replaced by the following:
\begin{align}
\label{eq:chi1}
\chi^{\alpha}_1= p_{1/2} \; \frac{(\mu^{\alpha}_{\rm eff})^2}{3k_B} \frac{1}{T-\Theta_{\alpha}} \;,
\end{align}
where $\Theta_{\alpha}$ is the paramagnetic Curie temperature.
The fitting results are shown in Figs.~\ref{fig:9}(e,f); nice agreements with experiments
at the whole temperature region can be obtained with $\Theta_{ab}=17$ K and $\Theta_{c}=6$ K.

Another thing we want to mention here is that one can also deduce the trigonal crystal field from the
$g$-factor anisotropy of $\widetilde{S}=1/2$ doublet, which are given by wave functions Eq.~\ref{eq:wf} as:
\begin{align}
g_{ab} &=4\sqrt{3}\mathcal{C}_1\mathcal{C}_3+4\mathcal{C}_2^2
-3\sqrt{2}\kappa\mathcal{C}_2\mathcal{C}_3 \; ,
\notag \\
g_c &=(6+3\kappa)\mathcal{C}_1^2+2\mathcal{C}_2^2
-(2+3\kappa)\mathcal{C}_3^2 \; .
\label{eq:g}
\end{align}
At cubic limit, we have $g_{ab}=g_c$. The anisotropy of the $g$-factors incorporates the information of the crystal fields
as shown in Fig.~\ref{fig:9}(g). For positive $\Delta$, we have $g_{ab} > g_c$, and hence $\chi_{ab}>\chi_c$, corresponding to the case of
Na$_3$Co$_2$SbO$_6$.

\begin{figure}
\begin{center}
\includegraphics[width=12.6cm]{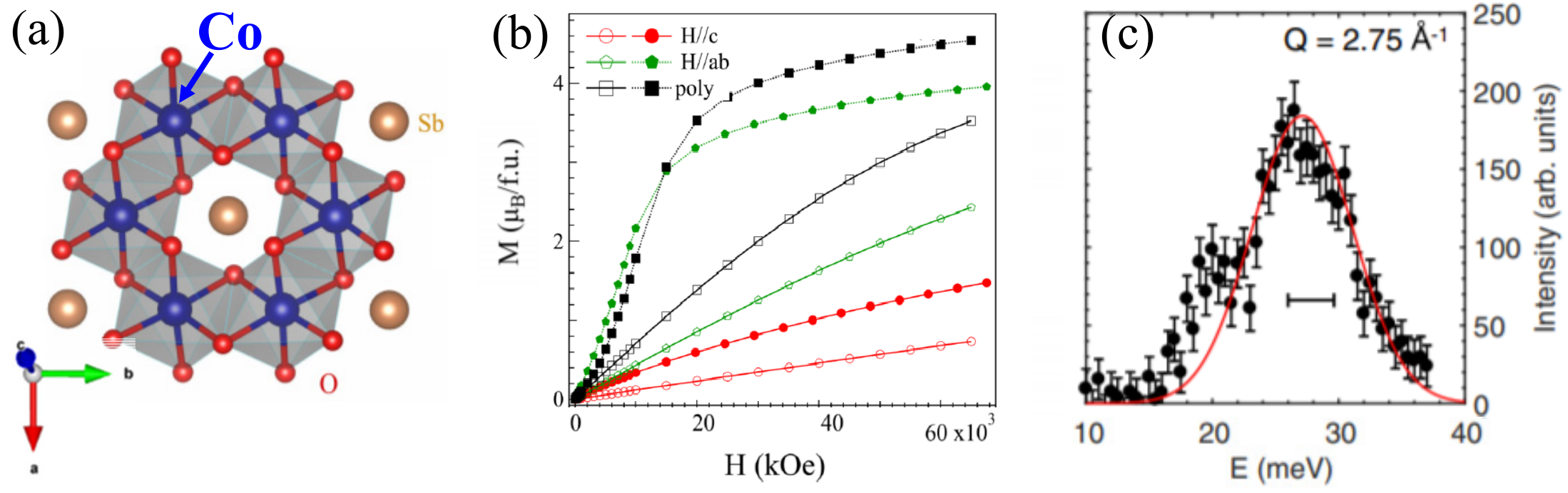}
\caption[]{(a) The honeycomb arrangement of Co ions in $ab$ plane
and (b) field dependence of magnetization of single crystals and polycrystals at 2 K (solid) and 30 K (open)
of Na$_3$Co$_2$SbO$_6$ from Ref.~\refcite{Yan19}.
(c) Constant-$Q$ scans of the spin-orbit excitations between the $1/2$ and $3/2$ manifolds in Na$_3$Co$_2$SbO$_6$,
 taken from Ref.~\refcite{Son20}. The excitation energy is at 27 meV.}
\label{fig:10}
\end{center}
\end{figure}

It is worth to comment on the positive sign of $\Delta>0$ in Na$_3$Co$_2$SbO$_6$.
Within a point-charge model when only the contribution from O$_6$ octahedron is considered,
one would find a negative $\Delta<0$ instead, since the octahedron is compressed along the $Z$-axis\cite{Yan19}.
However, the non-cubic Madelung potential of distant ions is neglected in this approximation.
In Na$_3$Co$_2$SbO$_6$, we think that $\Delta>0$ is due to a positive contribution of the
high-valence Sb$^{5+}$ ions residing within the $ab$-plane, see Fig.~\ref{fig:10}(a).
A $Z$-axis compression would give rise to the negative contribution of the oxygen octahedra and
compensate the positive $\Delta$ contributed by Sb$^{5+}$ ions, reducing thereby a total value of the trigonal field $\Delta$.

With $\Delta=38$ meV and $\lambda=28$ meV obtained above, we get $g_{ab}\simeq$~4.6 and $g_c\simeq$~3 for Na$_3$Co$_2$SbO$_6$, as shown in Fig.~\ref{fig:9}(g).
This gives the in-plane saturated magnetic moment $M_{ab}=g_{ab}\widetilde{S}=2.3\mu_B$, which is in agreement with the experimental value\cite{Yan19}.
In addition, one can also get that the spin-orbit exciton mode is located roughly $\sim$ 30 meV, and this is also consistent with the experiment as shown in Fig.~\ref{fig:10}(c).

\subsection{Mapping Na$_3$Co$_2$SbO$_6$ onto the phase diagram}
\label{sec:V}
By now, the $\Delta/\lambda \sim 1.36$ ratio is established in Na$_3$Co$_2$SbO$_6$.
Regarding the $U/\Delta_{pd}$ ratio, we know $U/\Delta_{pd} \sim 2 - 3$ is reasonable in cobaltates as mentioned above.
For Na$_3$Co$_2$SbO$_6$, we believe it is close to the phase boundary between zz1 and FM/\!/$ab$ phase.
This assumption is quite reasonable since the zigzag order gives way to fully polarized state at very small magnetic fields\cite{Vic07,Yan19}, as shown in Fig.~\ref{fig:10}(b).
In addition, a sister compound Li$_3$Co$_2$SbO$_6$ has $ab$-plane FM order\cite{Str19,Viv20} (most likely due to smaller Co-O-Co bond angle, $91^{\circ}$ versus $93^{\circ}$, slightly enhancing the FM $J$ value and thus stabilizing the FM order). These facts imply that zz1 and FM/\!/$ab$ states are indeed closely competing in Na$_3$Co$_2$SbO$_6$.
However, this is still not enough to quantify the $U/\Delta_{pd}$ ratio. From Figs.~\ref{fig:7} (a-f), it is clear that the  phase boundary between zz1 and FM/\!/$ab$ phase
varies for different $J_3$. This suggests that $J_3$ has to be fixed first before evaluating $U/\Delta_{pd}$ ratio.

In fact, the choice of $J_3$ can also be dictated by the close proximity of zz1 and FM/\!/$ab$ states in Na$_3$Co$_2$SbO$_6$.
The classical energies of these two states differ by:
\begin{align}
E_{\rm {FM/\!/}ab}-E_{\rm {zz1}}=\tfrac{1}{4}(J-\Gamma+3J_3) \;,
\end{align}
which is $ \sim 0 $ in Na$_3$Co$_2$SbO$_6$ and gives a rough idea of $J_3\sim -J/3$ (ignoring small $\Gamma$).
In the parameter space with $\Delta/\lambda \sim 1.36$ and $U/\Delta_{pd} \sim 2 - 3$, we have $|J|<0.8 t^2/U$, and thus $J_3 \sim -J/3 < 0.3 t^2/U$  can be estimated.

\begin{figure}
\begin{center}
\includegraphics[width=12.6cm]{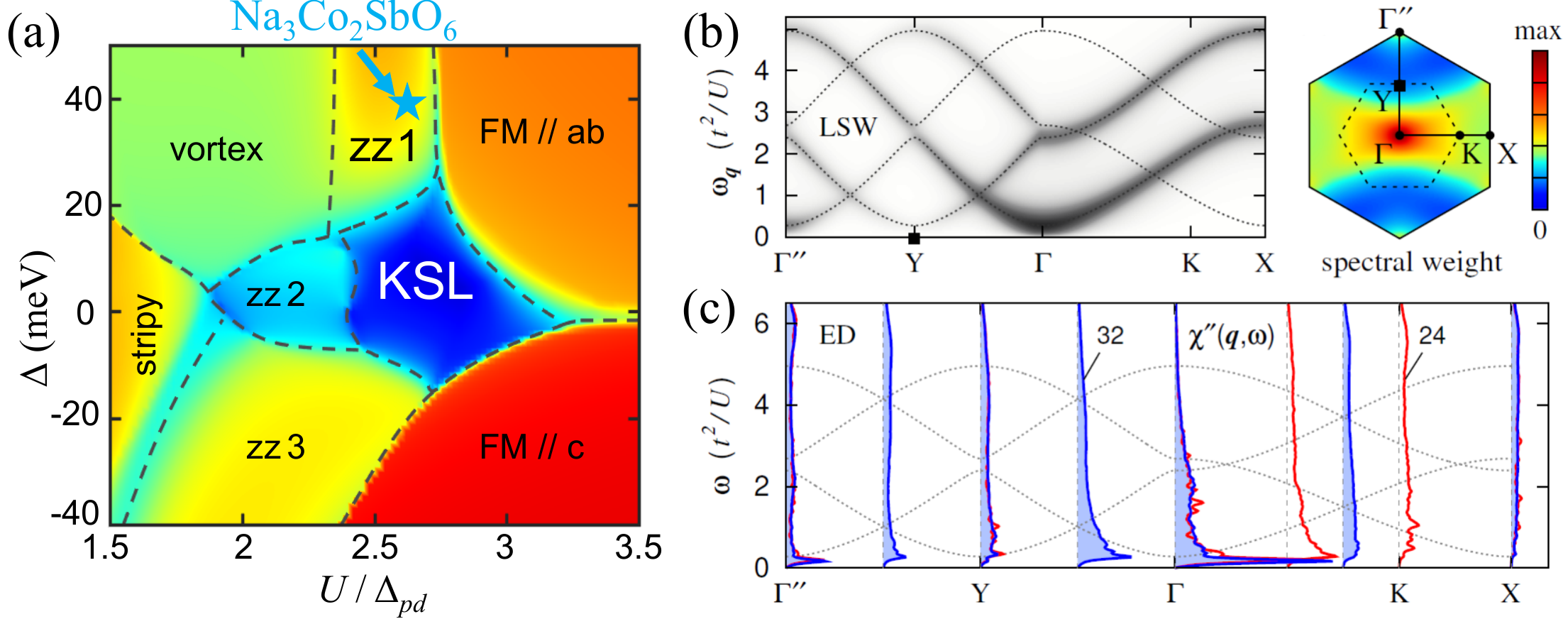}
\caption[]{(a) Rough position of  Na$_3$Co$_2$SbO$_6$ in the phase diagram with $J_3=0.15 t^2/U$ is indicated by the star.
Spin excitation spectrum expected in Na$_3$Co$_2$SbO$_6$
(b) from linear spin wave (LSW) theory and (c) ED results for hexagonal 24- and 32-site clusters
with $K=-3.6$, $J=-0.5$, $\Gamma=-0.1$, $\Gamma'=0.6$ and $J_3=0.15$ (in units of $t^2/U$).
The intensity is largest around $\Gamma$, i.e. away from the Bragg point Y.
Figure is taken from Ref.~\refcite{Liu20}.}
\label{fig:11}
\end{center}
\end{figure}

As an example, by taking the phase diagram with $J_3=0.15t^2/U\simeq 0.04|K|$, Na$_3$Co$_2$SbO$_6$ can be located at $U/\Delta_{pd} \sim 2.5$-2.7 and $\Delta/\lambda\sim 1.36$, see Fig.~\ref{fig:11}(a).
In this parameter area, the exchange couplings are $K\simeq -3.6\;t^2/U$, $J/|K|\sim -0.14$, $\Gamma/|K|\sim -0.03$, and $\Gamma'/|K|\sim 0.16$.
The small values of $J,\Gamma,\Gamma'$ imply the proximity to the Kitaev model, explaining a strong reduction of the ordered moments in this material from the saturated values\cite{Yan19}.
We can evaluate $\Theta$ values using the obtained theoretical exchange constants and $J_3=0.15 t^2/U$:
\begin{align}
\Theta_{ab}=-\tfrac{3}{4}\left [ J+J_3 + \tfrac{1}{3}K-\tfrac{1}{3}(\Gamma+2\Gamma') \right] \simeq 1.4 \;(t^2/U),  \notag \\
\Theta_{c}=-\tfrac{3}{4}\left [ J+J_3 + \tfrac{1}{3}K+\tfrac{2}{3}(\Gamma+2\Gamma') \right] \simeq 0.6 \;(t^2/U).
\label{eq:curie}
\end{align}
Curiously enough, this gives the $\Theta$-anisotropy close to what we get from the susceptibility fits.
Besides, this comparison also suggests the energy scale of $t^2/U\sim 1$~meV, setting thereby the magnon bandwidth of the order of $10$~meV. The relative smallness of $t^2/U$ is due to large $U$ and more localized nature of $3d$ orbitals.

As suggested by Fig.~\ref{fig:11}(a), Na$_3$Co$_2$SbO$_6$ is located at just $\sim 20$~meV ``distance'' from the KSL phase.
At this point, the relative smallness of SOC for 3$d$ Co ions comes as a great advantage:
on one hand it is strong enough to form the pseudospin moments, on the other hand it makes the lattice
manipulation of the $\widetilde{S}=1/2$ wave functions (and hence magnetism) far easier than in iridates\cite{Liu19}.

A question of experimental interest is how to drive Na$_3$Co$_2$SbO$_6$ into the KSL phase.
A reduction of the trigonal field by $\sim 20$~meV (compression along $Z$-axis) by means of strain or pressure
control seems feasible on experimental side, given that $\Delta$ variations within a window of $\sim 70$~meV were achieved by strain control in a cobalt oxide\cite{Csi05}.
Thus, monitoring the magnetic behavior of Na$_3$Co$_2$SbO$_6$ under uniaxial pressure would be very interesting.
Note that by compressing the materials along $Z$-axis, the distances between the in-plane Co$^{2+}$ ions should be enhanced  and thus effectively reduce the
longer range Heisenberg interactions. Also, NN FM $J<0$ will be suppressed,
since the exchange bonding angles between NN Co$^{2+}$ ions will be further deviated from 90$^{\circ}$ by a compression along $Z$-axis.
All in all, the lattice engineering seems to be a promising way to realize the KSL phase in cobaltates.

\subsection{Magnetic spectrum of Na$_3$Co$_2$SbO$_6$}
\label{sec:ncso}
Using the obtained exchange parameters $K=-3.6$, $J=-0.5$, $\Gamma=-0.1$, $\Gamma'=0.6$ in units of $t^2/U$ and add $J_3=0.15$ by hand,
one can calculate the expected spin excitations in Na$_3$Co$_2$SbO$_6$ with linear spin wave (LSW) theory as shown in Fig.~\ref{fig:11}(b).
There is a small gap $\sim 0.3~t^2/U$ at $\Gamma$ point and the intensity is anisotropic in momentum space.
Even with the zigzag in-plane magnetic order, the magnon spectral weight is condensed near $\Gamma$ point instead of
the Bragg point $Y$ due to the large FM Kitaev interaction. 

To account for the quantum effect, we have performed ED calculations of dynamical spin susceptibility.
Compared with the linear spin wave theory result, the ED results in Fig.~\ref{fig:11} (c) show that, as a consequence of the dominant Kitaev coupling, magnons are strongly renormalized and only survive at low energies, and a broad continuum of excitations\cite{Win17a,Goh17} as in $\alpha$-RuCl$_3$\cite{Ban18,San15} emerges. Neutron scattering experiments on Na$_3$Co$_2$SbO$_6$ are desired to verify these predictions.

Here, we want to emphasize an important aspect that one has to keep in mind while comparing the above ED results with the experimental data. Namely, the cluster ground state is fully symmetric and contains three degenerate zigzag directions with equal weights for the hexagonal clusters. The dynamical spin susceptibility obtained by ED method in Fig.~\ref{fig:11} (c) contains contributions from all these zigzag patterns.
On the other hand, the intensities calculated using the LSW theory Fig.~\ref{fig:11} (b) correspond to a single-domain crystal with one particular zigzag pattern.

\begin{figure}
\begin{center}
\includegraphics[width=12.6cm]{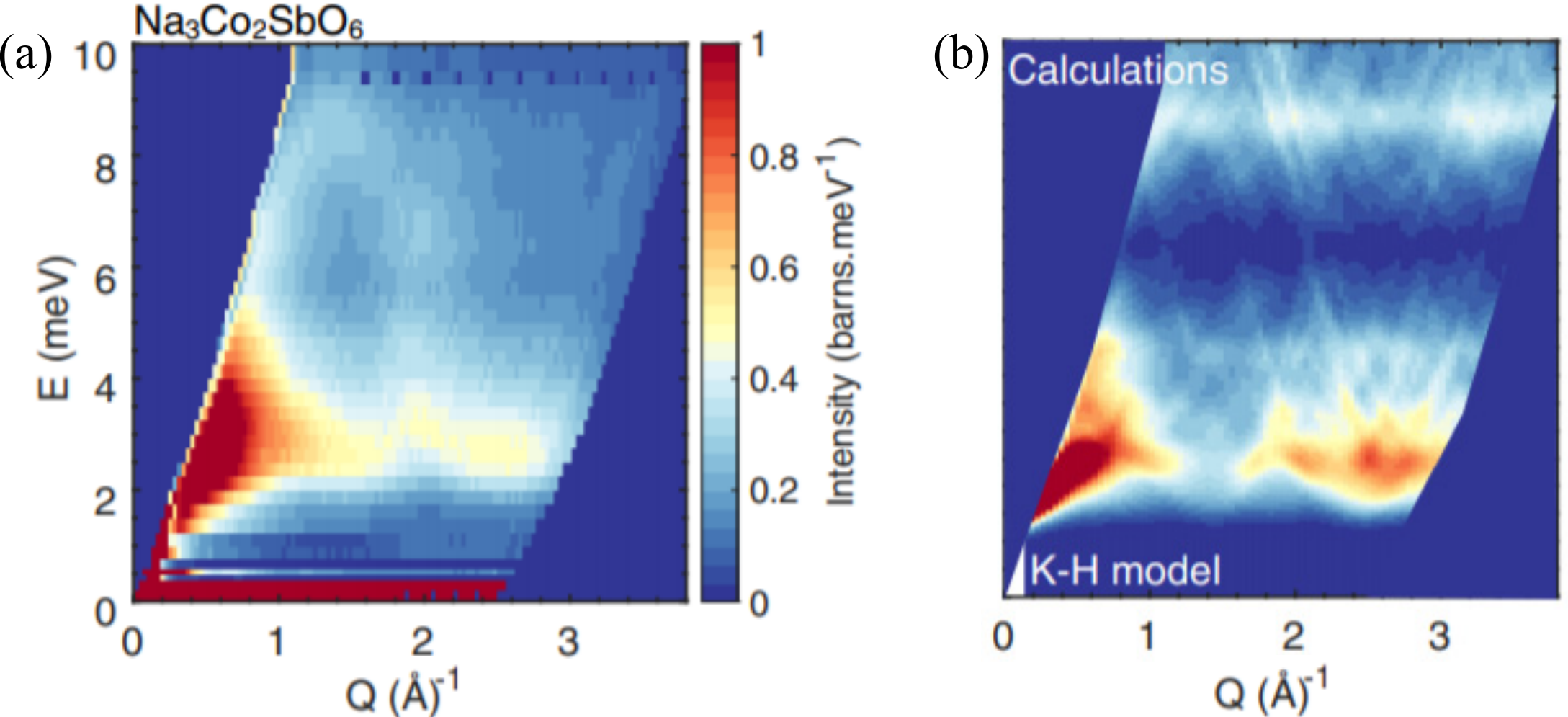}
\caption[]{(a) Dynamic structure factor measured at $T = 1.5 K$ and (b) spin-wave
calculations using Eq.~\ref{eq:H} with $(K, J, \Gamma, \Gamma',J_3)=(-9, -2, 0.3,-0.8, 0.8)$ meV in Na$_3$Co$_2$SbO$_6$.
Figure is taken from Ref.~\refcite{Son20}. }
\label{fig:12}
\end{center}
\end{figure}

Recently, inelastic neutron scattering measurement has been performed on the  polycrystalline sample
and the result is presented in Fig.~\ref{fig:12}(a). The bandwidth of the magnon spectrum is
within the order of 10 meV which is well below the spin-orbit exciton mode. As expected, the intensity
is condensed around $q=0$ point which is consistent with the above theoretical prediction, suggesting the presence of strong Kitaev coupling.
With the set of fitting parameters $(K, J, \Gamma, \Gamma',J_3)=(-9, -2, 0.3,-0.8, 0.8)$ meV, the experimental data has been roughly
reproduced\cite{Son20}, as shown in Fig.~\ref{fig:12}(b).

The experimental fitting suggests dominant FM Kitaev interaction $K<0$, sizable FM Heisenberg interaction $J<0$ and rather weak $\Gamma$ term.
These are consistent with the theoretical estimation $(K, J, \Gamma, \Gamma',J_3)=(-3.6, -0.5, -0.1,0.6, 0.15)~
t^2/U$. However, the sign of $\Gamma'$ term from the experimental fitting suggests negative $\Delta<0$, which is opposite from
the theoretical prediction. It is clear  from the above discussions, that the sign of $\Delta$ is crucial to the magnetic properties
and thus needs to be clarified in the future.
In particular, experimental studies on single crystal Na$_3$Co$_2$SbO$_6$ samples are highly desired.

\subsection{Na$_2$Co$_2$TeO$_6$}
\label{sec:ncto}
Recently, another cobalt compound Na$_2$Co$_2$TeO$_6$ has attracted research interest\cite{Son20,Che20,Lin20,Kim20,Vic07,Lef16,Ber17,Yao20,Hon21}.
Neutron diffraction studies\cite{Lef16,Ber17} have found long range zz1 order below $T_N\sim27$ K in Na$_2$Co$_2$TeO$_6$.
Since we can find the corresponding magnetic order in the phase diagram, it may be possible to map this material onto the phase diagram
and verify the potential of realizing the KSL phase in Na$_2$Co$_2$TeO$_6$.

\begin{figure}
\begin{center}
\includegraphics[width=12.6cm]{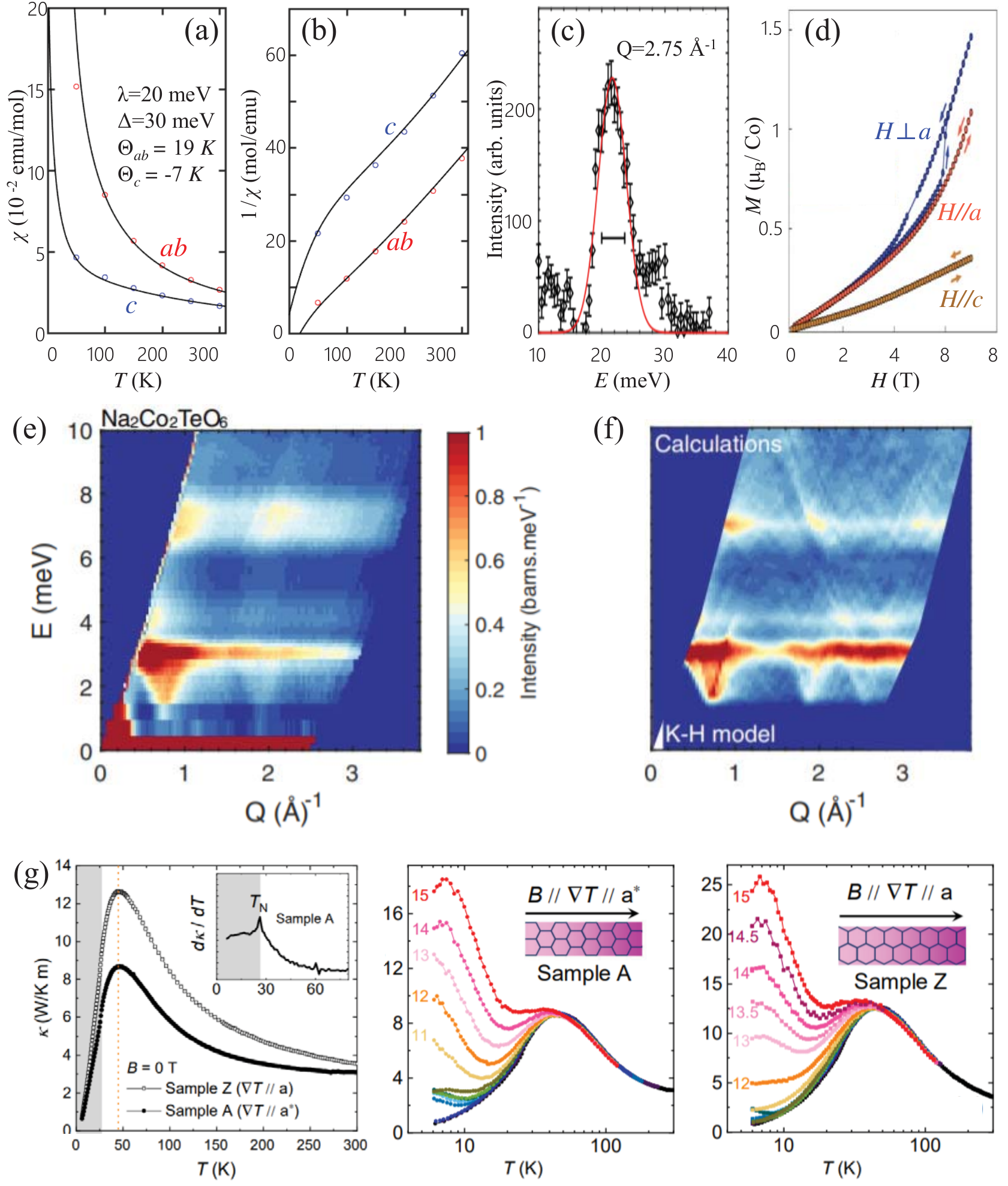}
\caption[]{Temperature dependence of (a) magnetic susceptibility
 and (b) its inverse in Na$_2$Co$_2$TeO$_6$. Open circles represent the experimental data extracted from Ref.~\refcite{Yao20}, and solid lines are the fits using single-ion
using $\chi^{ab}_0=1.4\times10^{-3}$ emu/mol, $\chi^c_0=-0.9\times 10^{-3}$ emu/mol and $\kappa=0.96$.
(c) Constant-$Q$ scans of the spin-orbit excitations between the $1/2$ and $3/2$ manifolds in Na$_2$Co$_2$TeO$_6$,
 taken from Ref.~\refcite{Son20}. The excitation energy is at 21.6 meV.
(d) Field dependence of magnetization of Na$_2$Co$_2$TeO$_6$ at 2 K taken from Ref.~\refcite{Yao20}.
(e) Dynamic structure factor measured at $T = 1.5 K$ and (f) spin-wave
calculations with $(K, J, \Gamma, \Gamma',J_2,J_3)=(-9, -0.1, 1.8, 0.3, 0.3, 0.9)$ meV in Na$_2$Co$_2$TeO$_6$ from Ref.~\refcite{Son20}.
(g) Temperature dependence of the thermal conductivity without and with in-plane magnetic field taken from Ref.~\refcite{Hon21}.
}
\label{fig:13}
\end{center}
\end{figure}

Following the previous steps, we show the paramagnetic susceptibility fits of the experimental data from Ref.~\refcite{Yao20}
in Figs.~\ref{fig:13}(a,b). $\chi_{ab}>\chi_c$ indicates positive $\Delta$ as in Na$_3$Co$_2$SbO$_6$.
Rather fair agreements can be achieved with $\lambda=20$ meV and $\Delta=30$ meV. This gives the spin-orbit exciton $1/2\rightarrow 3/2$ mode at $\sim21.6$ meV which is also quite consistent with the neutron data\cite{Son20} as shown in Fig.~\ref{fig:13}(c).
However, the obtained SOC constant of Na$_2$Co$_2$TeO$_6$ ($\lambda \sim 20$ meV) is much smaller than that of other cobaltates ($\lambda \sim 27-28$ meV).

Regardless, with the obtained $\Delta/\lambda=1.5$ from the magnetic susceptibility fit and similar $U/\Delta_{pd}$ ratio as in Na$_3$Co$_2$SbO$_6$,
one can get that the exchange parameters are $(K,J,\Gamma,\Gamma')\sim(-3.5,-0.5,-0.1,0.7)~t^2/U$ in Na$_2$Co$_2$TeO$_6$,
which are very similar with those in Na$_3$Co$_2$SbO$_6$. This is quite reasonable since there are no much difference of $\Delta/\lambda$ ratio between them.
The more obvious difference relies on the magnitude of $J_3$, which seems to be stronger than that in Na$_3$Co$_2$SbO$_6$
given the fact that the critical field of fully polarized state is rather high in Na$_2$Co$_2$TeO$_6$, see Fig.~\ref{fig:13}(d).
Unfortunately, the rather strong $J_3$ may prevent the formation of KSL phase and supports the zigzag order instead, as shown in Fig.~\ref{fig:7}.
Nevertheless, experiments are still needed to quantify the exact values of the exchange parameters.

Inelastic neutron scattering measurements have also been performed on polycrystalline Na$_2$Co$_2$TeO$_6$ sample\cite{Son20}, see Fig.~\ref{fig:13}(e).
Compared with the experimental data of Na$_3$Co$_2$SbO$_6$ in Fig.~\ref{fig:12}(a), one immediate difference is that the intensity
is drifted away from $\Gamma$ point to finite $Q\simeq 0.75 {\rm \AA} ^{-1}$ as well as the larger energy gap.

As  shown in Fig.~\ref{fig:13}(f), the fitted exchange parameters are $(K, J, \Gamma, \Gamma',J_2,J_3)=(-9, -0.1, 1.8, 0.3, 0.3, 0.9)$ meV in Na$_2$Co$_2$TeO$_6$.
Similar with  Na$_3$Co$_2$SbO$_6$, the FM Kitaev coupling  is also dominant here substantiating the universality of Eq.~\ref{eq:H} in layered cobaltates.
The FM $J<0$ is suppressed by further neighbor AFM Heisenberg interactions $J_2$ and $J_3$.
The different magnetic spectrum between Na$_3$Co$_2$SbO$_6$ and Na$_2$Co$_2$TeO$_6$  arise  from the stronger FM fluctuation in Na$_3$Co$_2$SbO$_6$, which has been pointed out also in $\alpha$-RuCl$_3$\cite{Suz20}.

The field dependent thermal conductivity of single crystal Na$_2$Co$_2$TeO$_6$ has also been studied\cite{Hon21}. In analogy to
the prime KSL candidate $\alpha$-RuCl$_3$\cite{Hen18}, the thermal conductivity is greatly enhanced by magnetic fields and
resembles a peculiar double-peak structure when changing temperatures, see Fig.~\ref{fig:13}(g), supporting the conjecture that Na$_2$Co$_2$TeO$_6$ being a potential materialization of the Kitaev model.

Other than the experimental results presented above, different scenarios have also been proposed by other groups, such as a negligible Kitaev interaction $K\sim0$\cite{Lin20} or AFM $K>0$\cite{Kim20} in Na$_2$Co$_2$TeO$_6$ deduced from neutron scattering experiments.
Besides, Li and his collaborators have suggested the magnetic ground state of Na$_2$Co$_2$TeO$_6$ is beyond zz1 order
considering the weak but canonical ferrimagnetic behavior under magnetic field\cite{Yao20,Xia19} as shown in Fig.~\ref{fig:13}(d). A triple-$\textbf{q}$ order formed by the superposition of three zigzag order parameters is proposed instead\cite{Che20}.
All these controversy debates call for further studies of Na$_2$Co$_2$TeO$_6$ both on theoretical and experimental sides.

\subsection{CoTiO$_3$}
\label{sec:cto}

Except the zz1 magnetic order discussed above, another common magnetic ground state for honeycomb cobaltates is FM$//$ab order in the phase diagram.
For instance, CoTiO$_3$ exhibits in-plane FM order, with FM planes stacked antiferromagnetically
along the $Z$-axis below $T_N\sim38$ K.

\begin{figure}
\begin{center}
\includegraphics[width=12.6cm]{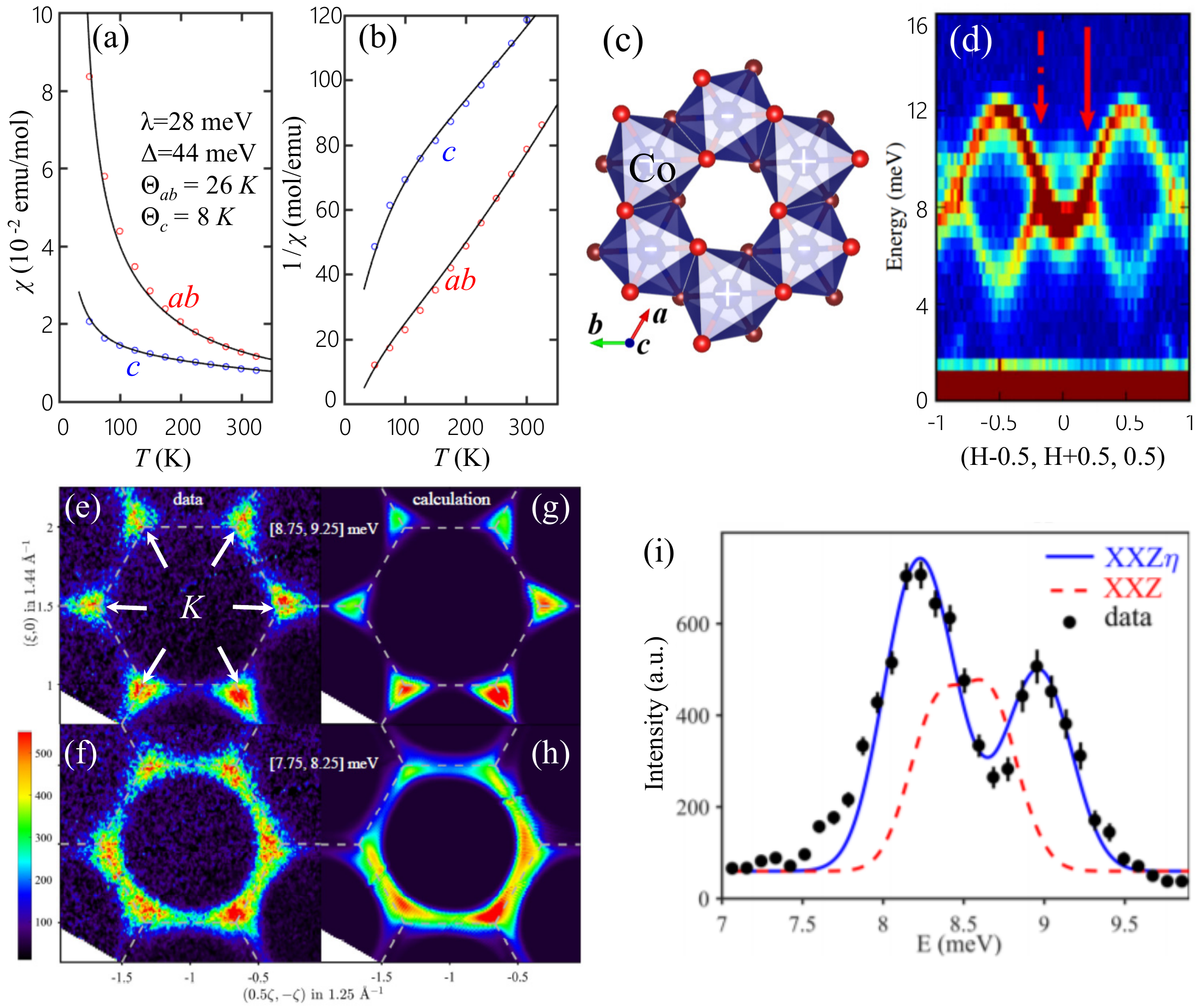}
\caption[]
{ Temperature dependence of (a) magnetic susceptibility
 and (b) its inverse in CoTiO$_3$ using $\chi^{ab}_0=3\times10^{-4}$ emu/mol, $\chi^c_0=8\times 10^{-4}$ emu/mol and $\kappa=0.96$.
 Open circles are experimental data from Ref.~\refcite{Bal17}.
 (c) Top view of the honeycomb plane of CoTiO$_3$, taken from Ref.~\refcite{Ell20}.
(d) Momentum and energy resolved neutron scattering intensity map of magnons in CoTiO$_3$,
two red arrows indicate the position of Dirac point where two linear magnon bands cross. Figure is taken from Ref.~\refcite{Yua20}.
(e-h) Experimental (left) and calculated (right) momentum intensity maps above and below the Dirac node energy, figure is taken from Ref.~\refcite{Ell20}.
(i) Averaged energy scan centred at K-points in (e), the experimental data shows clear two peaks feature which can be resolved by
$XXZ$ model with additional bond dependent anisotropy labeled by $\eta$. Figure is taken from Ref.~\refcite{Ell20}.}
\label{fig:14}
\end{center}
\end{figure}

Along the lines as in previous examples above, one can extract $\Delta=44$~meV and $\lambda=28$~meV by fitting the paramagnetic susceptibility data of CoTiO$_3$ from Ref.~\refcite{Bal17}, see Figs.~\ref{fig:14}(a,b).
The extracted values of $\Delta$ and $\lambda$ are quite consistent with the ones
obtained from neutron scattering experiments\cite{Yua20,Ell20}, which gives $\Delta/\lambda\sim 1.57$.
Together with $U/\Delta_{pd} \sim 2.5$-2.7 justified above,
one can estimate $(K,J,\Gamma,\Gamma')\sim(-3.5,-0.3,-0.14,0.6)~t^2/U$ in CoTiO$_3$ based on the theory.
On the other hand, $J_3$ is expected to be smaller in CoTiO$_3$ than in the above two cobaltates due to the lattice structure,
as there are no ions in the middle of the hexagon bridging the Co-Co neighbors as shown in Fig.~\ref{fig:14} (c).
Taking $J_3=0$ for instance, the Curie temperatures are evaluated as $\Theta_{ab}=1.51~t^2/U$ and $\Theta_c=0.64~t^2/U$
and the evaluated $\Theta$-anisotropy is close to what we get from the susceptibility fits.
The estimated exchange parameters locate CoTiO$_3$ in Fig.~\ref{fig:7}(a) rather close to the KSL phase.
Hence, further experimental studies of this material is desired.

As a matter of fact, the magnon dispersion of CoTiO$_3$ is of particular interest in the context of non-trivial magnon topology. 
A clear gapless Dirac cone has been revealed by the neutron scattering experiments\cite{Yua20,Ell20},
as shown in Fig.~\ref{fig:14}(d). A distinctive azimuthal modulation in the dynamical structure factor
around the linear touching Dirac points has been observed\cite{Ell20}, see Figs.~\ref{fig:14}(e,f), and this can be considered as the fingerprint of a topologically non-trivial magnon band structures\cite{Shi17}.
In Figs.~\ref{fig:14}(g,h), excellent agreement with the experimental data could be achieved by the $XXZ\eta$ model with $\eta$ representing the bond-dependent exchange anisotropies\cite{Ell20}. A more conclusive evidence of the presence of bond-dependent exchange interactions is shown in Fig.~\ref{fig:14}(i), where the two peaks feature of the average energy scan around the Dirac point can only be well explained by the the $XXZ\eta$ model\cite{Ell20}.
In addition, the high-resolution data in Ref.~\refcite{Ell20} presents
a small spectral gap of $\sim$1 meV at low energy which implies the existence of a quantum
order-by-disorder mechanism involving bond-dependent interactions.

In addition to the above discussed three materials, there are several other experiments that promote
the potential of realizing the Kitaev physics in $3d$ honeycomb cobaltates such as the magnetic
field induced spin-liquid-like behavior in BaCo$_2$(P$_{1-x}$V$_x$O$_4$)$_2$\cite{Zho18}
and nonmagnetic state in BaCo$_2$(AsO$_4$)$_2$\cite{Zho20}, signatures of Kitaev spin liquid physics in Li$_3$Co$_2$SbO$_6$ by neutron powder diffraction measurements, heat capacity, and magnetization studies\cite{Viv20}.
All these interesting results support importance of the further studies of honeycomb cobaltates.

To identify a promising KSL candidate material, it is indeed important to determine the exchange interactions.
One possibility is through analyzing the strengths of crystal field and SOC, which is emphasized in this review.
Another possibility is to extract the exchange parameters directly from thermodynamic, magnetic properties or magnetic excitation spectrum,
as we presented in the above Subsec. \ref{sec:ncso}-\ref{sec:cto}.
In addition to these, there are other proposals that are seemingly suitable to the cobaltates for future experimental investigations,
such as diluting the honeycomb magnets to remove the problematic molecular field \cite{Sar18,Sve78,Fur15} or distinct neutron-diffraction patterns of bond-dependent interactions measured in the paramagnetic phase\cite{Pad20}.

\section*{Summary}

As one goes from 5$d$ Ir to 4$d$ Ru and further to 3$d$ Co, magnetic $d$ orbitals become more localized. Therefore, non-Kitaev interactions related to the overlap between wave functions are expected to be smaller in 3$d$ systems, and
this should improve the conditions for realization of the NN-only interaction honeycomb model designed by Kitaev.
The idea of extending the research for KSL to $3d$ systems has already
led to a wealth of experiments, which in turn have provided valuable information and decisive evidences
on the universality of bond-dependent interactions in cobaltates.
Measurements of magnetic excitation spectra by inelastic neutron scattering, Raman
spectroscopy, electron spin resonance, NMR and THz spectroscopy are urgently needed
 to quantify the Hamiltonian parameters in various candidate materials.

Another important research direction to take will be  the lattice engineering of magnetism in cobaltates.
The trigonal crystal field, proposed as a key tuning parameter of the exchange Hamiltonian, can
decide the  proximity of a given material to the Kitaev spin liquid phase.
Monitoring the magnetic behavior of honeycomb cobaltates under uniaxial pressure
will be very useful   to verify this theoretical proposal.
In addition, the bond-angle control of the non-Kitaev exchange parameters can be realized through epitaxial strain.
In a broader context, doping of $d^7$ cobaltates, where magnetism is dominated by Kitaev-type interactions, would be highly interesting and may bring new surprises.

\section*{Acknowledgements}
We would like to thank G. Khaliullin, J. Chaloupka, R. Coldea, and Z. Z. Du for fruitful discussions.
The support by the European Research Council under Advanced Grant No. 669550 (Com4Com) is also acknowledged.

\end{document}